%% file: main.tex
\let\orienddocument\enddocument 
\documentclass[longauth]{aa}
\let\enddocument\orienddocument
\usepackage{lineno}
\usepackage{lastpage}
\usepackage{graphicx}
\usepackage[varg]{txfonts}
\usepackage{hyperref}
\hypersetup{
    colorlinks=true,
    linkcolor=blue,
    citecolor=blue,
    filecolor=blue,      
    urlcolor=blue,
    breaklinks=true,
}
\usepackage[normalem]{ulem} 
\usepackage{amssymb}

\usepackage[dvipsnames]{xcolor}

\newcommand{\trades}[0]{\textsc{TRADES}}
\newcommand{\photrades}[0]{\textsc{photoTRADES}}
\newcommand{\pyorbit}[0]{\textsc{PyORBIT}}
\newcommand{\pycheops}[0]{\textsc{pycheops}}
\newcommand{\pyde}[0]{\textsc{PyDE}}
\newcommand{\emcee}[0]{\textsc{emcee}}
\newcommand{\pipe}[0]{\textsc{PIPE}}
\newcommand{\pytransit}[0]{\textsc{PyTransit}}

\newcommand{\kepler}[0]{\textit{Kepler}}

\newcommand{\rmb}[0]{\ensuremath{\mathrm{b}}}
\newcommand{\rmc}[0]{\ensuremath{\mathrm{c}}}
\newcommand{\rmd}[0]{\ensuremath{\mathrm{d}}}

\newcommand{\mps}[0]{m\,s$^{-1}$}
\newcommand{\gcc}[0]{g\,\mbox{cm}$^{-3}$}

\usepackage{natbib}
\bibpunct{(}{)}{;}{a}{}{,} 

\begin{document}

   \title{
       Characterisation of the Warm-Jupiter TOI-1130 system with CHEOPS and photo-dynamical approach
       \thanks{This study uses CHEOPS data observed as part of the Guaranteed Time Observation (GTO) programmes
       CH\_PR00015, CH\_PR00031 and CH\_PR00053.
       Photometry of TESS, CHEOPS, and ASTEP+, and transit times prediction are only available in electronic form at the CDS via anonymous ftp to \url{cdsarc.u-strasbg.fr} (130.79.128.5) or via \url{http://cdsweb.u-strasbg.fr/cgi-bin/qcat?J/A+A/}.
       }
   }
   \titlerunning{TOI-1130 with CHEOPS.}


   \author{
        L.~Borsato\inst{\ref{inst1}} $^{\href{https://orcid.org/0000-0003-0066-9268}{\includegraphics[scale=0.04]{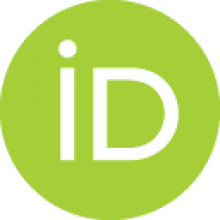}}}$ \and
        D.~Degen\inst{\ref{inst2}} $^{\href{https://orcid.org/0009-0008-1068-481X}{\includegraphics[scale=0.04]{orcid.png}}}$ \and
        A.~Leleu\inst{\ref{inst3},\ref{inst4}} $^{\href{https://orcid.org/0000-0003-2051-7974}{\includegraphics[scale=0.04]{orcid.png}}}$ \and
        M.J.~Hooton\inst{\ref{inst5}} $^{\href{https://orcid.org/0000-0003-0030-332X}{\includegraphics[scale=0.04]{orcid.png}}}$ \and
        J.A.~Egger\inst{\ref{inst4}} $^{\href{https://orcid.org/0000-0003-1628-4231}{\includegraphics[scale=0.04]{orcid.png}}}$ \and
        A.~Bekkelien\inst{\ref{inst3}} \and
        A.~Brandeker\inst{\ref{inst6}} $^{\href{https://orcid.org/0000-0002-7201-7536}{\includegraphics[scale=0.04]{orcid.png}}}$ \and
        A.~Collier Cameron\inst{\ref{inst7}} $^{\href{https://orcid.org/0000-0002-8863-7828}{\includegraphics[scale=0.04]{orcid.png}}}$ \and
        M.N.~Günther\inst{\ref{inst8}} $^{\href{https://orcid.org/0000-0002-3164-9086}{\includegraphics[scale=0.04]{orcid.png}}}$ \and
        V.~Nascimbeni\inst{\ref{inst1}} $^{\href{https://orcid.org/0000-0001-9770-1214}{\includegraphics[scale=0.04]{orcid.png}}}$ \and
        C.M.~Persson\inst{\ref{inst9}} \and
        A.~Bonfanti\inst{\ref{inst10}} $^{\href{https://orcid.org/0000-0002-1916-5935}{\includegraphics[scale=0.04]{orcid.png}}}$ \and
        T.G.~Wilson\inst{\ref{inst11}} $^{\href{https://orcid.org/0000-0001-8749-1962}{\includegraphics[scale=0.04]{orcid.png}}}$ \and
        A.C.M.~Correia\inst{\ref{inst12}} $^{\href{https://orcid.org/0000-0002-8946-8579}{\includegraphics[scale=0.04]{orcid.png}}}$ \and
        T.~Zingales\inst{\ref{inst13},\ref{inst1}} $^{\href{https://orcid.org/0000-0001-6880-5356}{\includegraphics[scale=0.04]{orcid.png}}}$ \and
        T.~Guillot\inst{\ref{inst14}} $^{\href{https://orcid.org/0000-0002-7188-8428}{\includegraphics[scale=0.04]{orcid.png}}}$ \and
        A.H.M.J.~Triaud\inst{\ref{inst15}} $^{\href{https://orcid.org/0000-0002-5510-8751}{\includegraphics[scale=0.04]{orcid.png}}}$ \and
        G.~Piotto\inst{\ref{inst1},\ref{inst13}} $^{\href{https://orcid.org/0000-0002-9937-6387}{\includegraphics[scale=0.04]{orcid.png}}}$ \and
        D.~Gandolfi\inst{\ref{inst16}} $^{\href{https://orcid.org/0000-0001-8627-9628}{\includegraphics[scale=0.04]{orcid.png}}}$ \and
        L.~Abe\inst{\ref{inst14}} $^{\href{https://orcid.org/0000-0002-0856-4527}{\includegraphics[scale=0.04]{orcid.png}}}$ \and
        Y.~Alibert\inst{\ref{inst17},\ref{inst4}} $^{\href{https://orcid.org/0000-0002-4644-8818}{\includegraphics[scale=0.04]{orcid.png}}}$ \and
        R.~Alonso\inst{\ref{inst18},\ref{inst19}} $^{\href{https://orcid.org/0000-0001-8462-8126}{\includegraphics[scale=0.04]{orcid.png}}}$ \and
        T.~Bárczy\inst{\ref{inst20}} $^{\href{https://orcid.org/0000-0002-7822-4413}{\includegraphics[scale=0.04]{orcid.png}}}$ \and
        D.~Barrado Navascues\inst{\ref{inst21}} $^{\href{https://orcid.org/0000-0002-5971-9242}{\includegraphics[scale=0.04]{orcid.png}}}$ \and
        S.C.C.~Barros\inst{\ref{inst22},\ref{inst23}} $^{\href{https://orcid.org/0000-0003-2434-3625}{\includegraphics[scale=0.04]{orcid.png}}}$ \and
        W.~Baumjohann\inst{\ref{inst10}} $^{\href{https://orcid.org/0000-0001-6271-0110}{\includegraphics[scale=0.04]{orcid.png}}}$ \and
        T.~Beck\inst{\ref{inst4}} \and
        P.~Bendjoya\inst{\ref{inst14}} $^{\href{https://orcid.org/0000-0002-4278-1437}{\includegraphics[scale=0.04]{orcid.png}}}$ \and
        W.~Benz\inst{\ref{inst4},\ref{inst17}} $^{\href{https://orcid.org/0000-0001-7896-6479}{\includegraphics[scale=0.04]{orcid.png}}}$ \and
        N.~Billot\inst{\ref{inst3}} $^{\href{https://orcid.org/0000-0003-3429-3836}{\includegraphics[scale=0.04]{orcid.png}}}$ \and
        C.~Broeg\inst{\ref{inst4},\ref{inst17}} $^{\href{https://orcid.org/0000-0001-5132-2614}{\includegraphics[scale=0.04]{orcid.png}}}$ \and
        M.-D.~Busch\inst{\ref{inst24}} \and
        Sz.~Csizmadia\inst{\ref{inst25}} $^{\href{https://orcid.org/0000-0001-6803-9698}{\includegraphics[scale=0.04]{orcid.png}}}$ \and
        P.E.~Cubillos\inst{\ref{inst10},\ref{inst26}} \and
        M.B.~Davies\inst{\ref{inst27}} $^{\href{https://orcid.org/0000-0001-6080-1190}{\includegraphics[scale=0.04]{orcid.png}}}$ \and
        M.~Deleuil\inst{\ref{inst28}} $^{\href{https://orcid.org/0000-0001-6036-0225}{\includegraphics[scale=0.04]{orcid.png}}}$ \and
        A.~Deline\inst{\ref{inst3}} \and
        L.~Delrez\inst{\ref{inst29},\ref{inst30},\ref{inst31}} $^{\href{https://orcid.org/0000-0001-6108-4808}{\includegraphics[scale=0.04]{orcid.png}}}$ \and
        O.D.S.~Demangeon\inst{\ref{inst22},\ref{inst23}} $^{\href{https://orcid.org/0000-0001-7918-0355}{\includegraphics[scale=0.04]{orcid.png}}}$ \and
        B.-O.~Demory\inst{\ref{inst17},\ref{inst4}} $^{\href{https://orcid.org/0000-0002-9355-5165}{\includegraphics[scale=0.04]{orcid.png}}}$ \and
        A.~Derekas\inst{\ref{inst32}} \and
        B.~Edwards\inst{\ref{inst33}} \and
        D.~Ehrenreich\inst{\ref{inst3},\ref{inst34}} $^{\href{https://orcid.org/0000-0001-9704-5405}{\includegraphics[scale=0.04]{orcid.png}}}$ \and
        A.~Erikson\inst{\ref{inst25}} \and
        A.~Fortier\inst{\ref{inst4},\ref{inst17}} $^{\href{https://orcid.org/0000-0001-8450-3374}{\includegraphics[scale=0.04]{orcid.png}}}$ \and
        L.~Fossati\inst{\ref{inst10}} $^{\href{https://orcid.org/0000-0003-4426-9530}{\includegraphics[scale=0.04]{orcid.png}}}$ \and
        M.~Fridlund\inst{\ref{inst35},\ref{inst9}} $^{\href{https://orcid.org/0000-0002-0855-8426}{\includegraphics[scale=0.04]{orcid.png}}}$ \and
        K.~Gazeas\inst{\ref{inst36}} \and
        M.~Gillon\inst{\ref{inst29}} $^{\href{https://orcid.org/0000-0003-1462-7739}{\includegraphics[scale=0.04]{orcid.png}}}$ \and
        M.~Güdel\inst{\ref{inst37}} \and
        A.~Heitzmann\inst{\ref{inst3}} $^{\href{https://orcid.org/0000-0002-8091-7526}{\includegraphics[scale=0.04]{orcid.png}}}$ \and
        Ch.~Helling\inst{\ref{inst10},\ref{inst38}} \and
        S.~Hoyer\inst{\ref{inst28}} $^{\href{https://orcid.org/0000-0003-3477-2466}{\includegraphics[scale=0.04]{orcid.png}}}$ \and
        K.G.~Isaak\inst{\ref{inst8}} $^{\href{https://orcid.org/0000-0001-8585-1717}{\includegraphics[scale=0.04]{orcid.png}}}$ \and
        L.L.~Kiss\inst{\ref{inst39},\ref{inst40}} \and
        J.~Korth\inst{\ref{inst41}} $^{\href{https://orcid.org/0000-0002-0076-6239}{\includegraphics[scale=0.04]{orcid.png}}}$ \and
        K.W.F.~Lam\inst{\ref{inst25}} $^{\href{https://orcid.org/0000-0002-9910-6088}{\includegraphics[scale=0.04]{orcid.png}}}$ \and
        J.~Laskar\inst{\ref{inst42}} $^{\href{https://orcid.org/0000-0003-2634-789X}{\includegraphics[scale=0.04]{orcid.png}}}$ \and
        A.~Lecavelier des Etangs\inst{\ref{inst43}} $^{\href{https://orcid.org/0000-0002-5637-5253}{\includegraphics[scale=0.04]{orcid.png}}}$ \and
        M.~Lendl\inst{\ref{inst3}} $^{\href{https://orcid.org/0000-0001-9699-1459}{\includegraphics[scale=0.04]{orcid.png}}}$ \and
        D.~Magrin\inst{\ref{inst1}} $^{\href{https://orcid.org/0000-0003-0312-313X}{\includegraphics[scale=0.04]{orcid.png}}}$ \and
        L.~Marafatto\inst{\ref{inst1}} $^{\href{https://orcid.org/0000-0002-8822-6834}{\includegraphics[scale=0.04]{orcid.png}}}$ \and
        P.F.L.~Maxted\inst{\ref{inst44}} $^{\href{https://orcid.org/0000-0003-3794-1317}{\includegraphics[scale=0.04]{orcid.png}}}$ \and
        M.~Mecina\inst{\ref{inst37}} $^{\href{https://orcid.org/0000-0002-3258-7526}{\includegraphics[scale=0.04]{orcid.png}}}$ \and
        D.~Mékarnia\inst{\ref{inst14}} $^{\href{https://orcid.org/0000-0001-5000-7292}{\includegraphics[scale=0.04]{orcid.png}}}$ \and
        C.~Mordasini\inst{\ref{inst4},\ref{inst17}} \and
        D.~Mura\inst{\ref{inst45},\ref{inst46}} $^{\href{https://orcid.org/0000-0002-9576-1106}{\includegraphics[scale=0.04]{orcid.png}}}$ \and
        G.~Olofsson\inst{\ref{inst6}} $^{\href{https://orcid.org/0000-0003-3747-7120}{\includegraphics[scale=0.04]{orcid.png}}}$ \and
        R.~Ottensamer\inst{\ref{inst37}} \and
        I.~Pagano\inst{\ref{inst47}} $^{\href{https://orcid.org/0000-0001-9573-4928}{\includegraphics[scale=0.04]{orcid.png}}}$ \and
        E.~Pallé\inst{\ref{inst18},\ref{inst19}} $^{\href{https://orcid.org/0000-0003-0987-1593}{\includegraphics[scale=0.04]{orcid.png}}}$ \and
        G.~Peter\inst{\ref{inst48}} $^{\href{https://orcid.org/0000-0001-6101-2513}{\includegraphics[scale=0.04]{orcid.png}}}$ \and
        D.~Pollacco\inst{\ref{inst11}} \and
        D.~Queloz\inst{\ref{inst2},\ref{inst5}} $^{\href{https://orcid.org/0000-0002-3012-0316}{\includegraphics[scale=0.04]{orcid.png}}}$ \and
        R.~Ragazzoni\inst{\ref{inst1},\ref{inst13}} $^{\href{https://orcid.org/0000-0002-7697-5555}{\includegraphics[scale=0.04]{orcid.png}}}$ \and
        N.~Rando\inst{\ref{inst8}} \and
        F.~Ratti\inst{\ref{inst8}} \and
        H.~Rauer\inst{\ref{inst25},\ref{inst49}} $^{\href{https://orcid.org/0000-0002-6510-1828}{\includegraphics[scale=0.04]{orcid.png}}}$ \and
        I.~Ribas\inst{\ref{inst50},\ref{inst51}} $^{\href{https://orcid.org/0000-0002-6689-0312}{\includegraphics[scale=0.04]{orcid.png}}}$ \and
        S.~Salmon\inst{\ref{inst3}} $^{\href{https://orcid.org/0000-0002-1714-3513}{\includegraphics[scale=0.04]{orcid.png}}}$ \and
        N.C.~Santos\inst{\ref{inst22},\ref{inst23}} $^{\href{https://orcid.org/0000-0003-4422-2919}{\includegraphics[scale=0.04]{orcid.png}}}$ \and
        G.~Scandariato\inst{\ref{inst47}} $^{\href{https://orcid.org/0000-0003-2029-0626}{\includegraphics[scale=0.04]{orcid.png}}}$ \and
        D.~Ségransan\inst{\ref{inst3}} $^{\href{https://orcid.org/0000-0003-2355-8034}{\includegraphics[scale=0.04]{orcid.png}}}$ \and
        A.E.~Simon\inst{\ref{inst4},\ref{inst17}} $^{\href{https://orcid.org/0000-0001-9773-2600}{\includegraphics[scale=0.04]{orcid.png}}}$ \and
        A.M.S.~Smith\inst{\ref{inst25}} $^{\href{https://orcid.org/0000-0002-2386-4341}{\includegraphics[scale=0.04]{orcid.png}}}$ \and
        S.G.~Sousa\inst{\ref{inst22}} $^{\href{https://orcid.org/0000-0001-9047-2965}{\includegraphics[scale=0.04]{orcid.png}}}$ \and
        M.~Stalport\inst{\ref{inst30},\ref{inst29}} \and
        O.~Suarez\inst{\ref{inst14}} $^{\href{https://orcid.org/0000-0002-3503-3617}{\includegraphics[scale=0.04]{orcid.png}}}$ \and
        S.~Sulis\inst{\ref{inst28}} $^{\href{https://orcid.org/0000-0001-8783-526X}{\includegraphics[scale=0.04]{orcid.png}}}$ \and
        Gy.M.~Szabó\inst{\ref{inst32},\ref{inst52}} $^{\href{https://orcid.org/0000-0002-0606-7930}{\includegraphics[scale=0.04]{orcid.png}}}$ \and
        S.~Udry\inst{\ref{inst3}} $^{\href{https://orcid.org/0000-0001-7576-6236}{\includegraphics[scale=0.04]{orcid.png}}}$ \and
        V.~Van Grootel\inst{\ref{inst30}} $^{\href{https://orcid.org/0000-0003-2144-4316}{\includegraphics[scale=0.04]{orcid.png}}}$ \and
        J.~Venturini\inst{\ref{inst3}} $^{\href{https://orcid.org/0000-0001-9527-2903}{\includegraphics[scale=0.04]{orcid.png}}}$ \and
        E.~Villaver\inst{\ref{inst18},\ref{inst19}} \and
        N.A.~Walton\inst{\ref{inst53}} $^{\href{https://orcid.org/0000-0003-3983-8778}{\includegraphics[scale=0.04]{orcid.png}}}$ \and
        D.~Wolter\inst{\ref{inst25}}
    }
    \authorrunning{Borsato et al.}

   \institute{
        \label{inst1} INAF, Osservatorio Astronomico di Padova, Vicolo dell'Osservatorio 5, 35122 Padova, Italy \and
        \label{inst2} ETH Zurich, Department of Physics, Wolfgang-Pauli-Strasse 2, CH-8093 Zurich, Switzerland \and
        \label{inst3} Observatoire astronomique de l'Universit\'{e} de Gen\`{e}ve, Chemin Pegasi 51, 1290 Versoix, Switzerland \and
        \label{inst4} Weltraumforschung und Planetologie, Physikalisches Institut, University of Bern, Gesellschaftsstrasse 6, 3012 Bern, Switzerland \and
        \label{inst5} Cavendish Laboratory, JJ Thomson Avenue, Cambridge CB3 0HE, UK \and
        \label{inst6} Department of Astronomy, Stockholm University, AlbaNova University Center, 10691 Stockholm, Sweden \and
        \label{inst7} Centre for Exoplanet Science, SUPA School of Physics and Astronomy, University of St Andrews, North Haugh, St Andrews KY16 9SS, UK \and
        \label{inst8} European Space Agency (ESA), European Space Research and Technology Centre (ESTEC), Keplerlaan 1, 2201 AZ Noordwijk, The Netherlands \and
        \label{inst9} Department of Space, Earth and Environment, Chalmers University of Technology, Onsala Space Observatory, 439 92 Onsala, Sweden \and
        \label{inst10} Space Research Institute, Austrian Academy of Sciences, Schmiedlstrasse 6, A-8042 Graz, Austria \and
        \label{inst11} Department of Physics, University of Warwick, Gibbet Hill Road, Coventry CV4 7AL, United Kingdom \and
        \label{inst12} CFisUC, Departamento de F\'{i}sica, Universidade de Coimbra, 3004-516 Coimbra, Portugal \and
        \label{inst13} Dipartimento di Fisica, Università degli Studi di Torino, via Pietro Giuria 1, I-10125, Torino, Italy \and
        \label{inst14} Dipartimento di Fisica e Astronomia "Galileo Galilei", Università degli Studi di Padova, Vicolo dell'Osservatorio 3, 35122 Padova, Italy \and
        \label{inst15} Laboratoire Lagrange, UMR7293, Observatoire de la C\^ote d'Azur, Universit\'e C\^ote d'Azur, Boulevard de l'Observatoire CS 34229 06304 Nice Cedex , France \and
        \label{inst16} School of Physics and Astronomy, University of Birmingham, Edgbaston, Birmingham B15 2TT, UK \and
        \label{inst17} Center for Space and Habitability, University of Bern, Gesellschaftsstrasse 6, 3012 Bern, Switzerland \and
        \label{inst18} Instituto de Astrofísica de Canarias, Vía Láctea s/n, 38200 La Laguna, Tenerife, Spain \and
        \label{inst19} Departamento de Astrofísica, Universidad de La Laguna, Astrofísico Francisco Sanchez s/n, 38206 La Laguna, Tenerife, Spain \and
        \label{inst20} Admatis, 5. Kand\'{o} Kálmán Street, 3534 Miskolc, Hungary \and
        \label{inst21} Depto. de Astrofísica, Centro de Astrobiología (CSIC-INTA), ESAC campus, 28692 Villanueva de la Cañada (Madrid), Spain \and
        \label{inst22} Instituto de Astrofisica e Ciencias do Espaco, Universidade do Porto, CAUP, Rua das Estrelas, 4150-762 Porto, Portugal \and
        \label{inst23} Departamento de Fisica e Astronomia, Faculdade de Ciencias, Universidade do Porto, Rua do Campo Alegre, 4169-007 Porto, Portugal \and
        \label{inst24} University of Bern, Sidlerstrasse 5, 3012 Bern \and
        \label{inst25} Institute of Planetary Research, German Aerospace Center (DLR), Rutherfordstrasse 2, 12489 Berlin, Germany \and
        \label{inst26} INAF, Osservatorio Astrofisico di Torino, Via Osservatorio, 20, I-10025 Pino Torinese To, Italy \and
        \label{inst27} Centre for Mathematical Sciences, Lund University, Box 118, 221 00 Lund, Sweden \and
        \label{inst28} Aix Marseille Univ, CNRS, CNES, LAM, 38 rue Fr\'{e}d\'{e}ric Joliot-Curie, 13388 Marseille, France \and
        \label{inst29} Astrobiology Research Unit, Universit\'{e} de Li\`{e}ge, All\'{e}e du 6 Août 19C, B-4000 Li\`{e}ge, Belgium \and
        \label{inst30} Space sciences, Technologies and Astrophysics Research (STAR) Institute, Universit\'{e} de Li\`{e}ge, All\'{e}e du 6 Août 19C, 4000 Li\`{e}ge, Belgium \and
        \label{inst31} Institute of Astronomy, KU Leuven, Celestijnenlaan 200D, 3001 Leuven, Belgium \and
        \label{inst32} ELTE Gothard Astrophysical Observatory, 9700 Szombathely, Szent Imre h. u. 112, Hungary \and
        \label{inst33} SRON Netherlands Institute for Space Research, Niels Bohrweg 4, 2333 CA Leiden, Netherlands \and
        \label{inst34} Centre Vie dans l’Univers, Facult\'{e} des sciences, Universit\'{e} de Gen\`{e}ve, Quai Ernest-Ansermet 30, 1211 Gen\`{e}ve 4, Switzerland \and
        \label{inst35} Leiden Observatory, University of Leiden, PO Box 9513, 2300 RA Leiden, The Netherlands \and
        \label{inst36} National and Kapodistrian University of Athens, Department of Physics, University Campus, Zografos GR-157 84, Athens, Greece \and
        \label{inst37} Department of Astrophysics, University of Vienna, Türkenschanzstrasse 17, 1180 Vienna, Austria \and
        \label{inst38} Institute for Theoretical Physics and Computational Physics, Graz University of Technology, Petersgasse 16, 8010 Graz, Austria \and
        \label{inst39} Konkoly Observatory, Research Centre for Astronomy and Earth Sciences, 1121 Budapest, Konkoly Thege Mikl\'{o}s út 15-17, Hungary \and
        \label{inst40} ELTE E\"otv\"os Lor\'and University, Institute of Physics, P\'azm\'any P\'eter s\'et\'any 1/A, 1117 Budapest, Hungary \and
        \label{inst41} Lund Observatory, Division of Astrophysics, Department of Physics, Lund University, Box 118, 22100 Lund, Sweden \and
        \label{inst42} IMCCE, UMR8028 CNRS, Observatoire de Paris, PSL Univ., Sorbonne Univ., 77 av. Denfert-Rochereau, 75014 Paris, France \and
        \label{inst43} Institut d'astrophysique de Paris, UMR7095 CNRS, Universit\'{e} Pierre \& Marie Curie, 98bis blvd. Arago, 75014 Paris, France \and
        \label{inst44} Astrophysics Group, Lennard Jones Building, Keele University, Staffordshire, ST5 5BG, United Kingdom \and
        \label{inst45} Istituto di Scienze Polari del Consiglio Nazionale delle Ricerche (CNR-ISP), via Torino 155, 30172 Venezia-Mestre, Italy \and
        \label{inst46} Programma Nazionale di Ricerche in Antartide (PNRA), Lungotevere Grande Ammiraglio Thaon di Revel 76, 00196 Rome, Italy \and
        \label{inst47} INAF, Osservatorio Astrofisico di Catania, Via S. Sofia 78, 95123 Catania, Italy \and
        \label{inst48} Institute of Optical Sensor Systems, German Aerospace Center (DLR), Rutherfordstrasse 2, 12489 Berlin, Germany \and
        \label{inst49} Institut fuer Geologische Wissenschaften, Freie Universitaet Berlin, Maltheserstrasse 74-100,12249 Berlin, Germany \and
        \label{inst50} Institut de Ciencies de l'Espai (ICE, CSIC), Campus UAB, Can Magrans s/n, 08193 Bellaterra, Spain \and
        \label{inst51} Institut d'Estudis Espacials de Catalunya (IEEC), 08860 Castelldefels (Barcelona), Spain \and
        \label{inst52} HUN-REN-ELTE Exoplanet Research Group, Szent Imre h. u. 112., Szombathely, H-9700, Hungary \and
        \label{inst53} Institute of Astronomy, University of Cambridge, Madingley Road, Cambridge, CB3 0HA, United Kingdom
    }

   \date{Received June 03, 2024; accepted July 06, 2024}

 
  \abstract
   {
    Among the thousands of exoplanets discovered to date, 
    approximately a few hundred gas giants on short-period orbits 
    are classified as "lonely"
    and only a few are in a multi-planet system  
    with a smaller companion on a close orbit. 
    The processes that formed multi-planet systems hosting gas giants
    on close orbits are poorly understood, 
    and only a few examples of this kind of system have been observed and well characterised.
   }
   {
    Within the contest of multi-planet system hosting gas-giant on short orbits, 
    we characterise TOI-1130 system by measuring masses and orbital parameters.
    This is a 2-transiting planet system with 
    a Jupiter-like planet (c) on a 8.35 days orbit and 
    a Neptune-like planet (b) on an inner (4.07 days) orbit.
    Both planets show strong anti-correlated
    transit timing variations (TTVs). 
    Furthermore, radial velocity (RV) analysis showed
    an additional linear trend, a possible hint of 
    a non-transiting candidate planet on a far outer orbit.
    }
   {
    Since 2019, extensive transit and radial velocity observations of the TOI-1130
    have been acquired using TESS and various ground-based facilities.
    We present a new photo-dynamical analysis of all available transit and RV data, 
    with the addition of new CHEOPS and ASTEP+ data that achieve 
    the best precision to date on the planetary radii and masses
    and on the timings of each transit.
   }
   {
    We were able to model interior structure of planet b constraining the presence of a gaseous envelope of H/He, 
    while it was not possible to assess the possible water content.
    Furthermore, we analysed the resonant state of the two transiting planets, 
    and we found that they lie just outside the resonant region. 
    This could be the result of the tidal evolution that the system underwent.
    We obtained both masses of the planets with a precision less than $1.5\%$, 
    and radii with a precision of about $1\%$ and $3\%$ for planet b and c, respectively.
    }
   {}


   \maketitle
\nolinenumbers

\section{Introduction}

Among the more of 5\,500\footnote{NASA Exoplanet Archive at 2024-04-22.} confirmed exoplanets,
about 500 are classified as "lonely" hot Jupiters 
\citep[HJs,][]{Latham2011ApJ...732L..24L, Steffen2012PNAS..109.7982S,Huang2016ApJ...825...98H, Schlaufman2016ApJ...825...62S},
gas giants on short orbital periods
.
Of those only a few are part of multi-planet systems 
hosting smaller companions on close (inner) orbits, e.g. 
WASP-47 \citep{Hellier2012MNRAS.426..739H, Becker2015ApJ...812L..18B, Bryant2022AJ....163..197B, Nascimbeni2023A&A...673A..42N},
Kepler-730 \citep{Zhu2018RNAAS...2..160Z,Canas2019ApJ...870L..17C},
TOI-5398 \citep{Mantovan2022MNRAS.516.4432M,Mantovan2023arXiv231016888M},
and, the subject of this work, TOI-1130 \citep{Huang2020ApJ...892L...7H, Korth2023A&A...675A.115K}.
However the definition of HJ is not so strict, and it sometimes overlap with
warm Jupiter (WJ) exoplanets, gas giants with periods of $\sim8-200$~d \citep{Huang2016ApJ...825...98H}.
These WJs show different orbital configurations compared with HJs,
as \citet{Huang2016ApJ...825...98H} found that about $50\%$ of the \kepler{} sample
of WJs are in multi-planet systems.
This lead to infer that the formation and evolution processes \citep{Wu2018AJ....156...96W, Kley2019}
of WJ and HJ are different.
The characterisation of gas-giant systems will enable us to ascertain which migration process the system underwent:
disk-driven migration \citep{Lin1996Natur.380..606L, Baruteau2016SSRv..205...77B}
or high-eccentricity migration \citep[HEM,][]{Rasio1996Sci...274..954R,Chatterjee2008ApJ...686..580C,Nagasawa2008ApJ...678..498N}.
It has been suggested by \citet{Vick2019MNRAS.484.5645V, Vick2023ApJ...943L..13V} and \citet{Jackson2023AJ....165...82J}
that the main process to form HJs is the HEM,
while there is no clear hint for a dominant mechanism to form WJ systems \citep{Borsato2021MNRAS.506.3810B}.\par

We decided to observe TOI-1130 \citep{Huang2020ApJ...892L...7H, Korth2023A&A...675A.115K} 
a 2-transiting planet system hosting a gas giant TOI-1130\,c on 8~d period-orbit, and
a lower-mass planet, TOI-1130\, b, on a inner orbit ($P_\rmb\sim4$~d),
and a linear trend in the radial velocity data could be a hint 
of an additional candidate planet on far outer orbit.
We collected (see Sec.~\ref{sec:obs})
published transit and radial velocity (RV) data and
new transit observations with 
the CHaracterising ExOPlanet Satellite \citep[CHEOPS,][]{Benz2021ExA....51..109B}
and ASTEP+ \citep{astep+2022SPIE12182E..2OS},
and an additional sector of the Transiting Exoplanet Survey Satellite \citep[TESS;][]{Ricker2015JATIS...1a4003R}.
We updated the stellar parameters (Sec.~\ref{sec:star}) and
we improved the radii measurement, masses and the architecture of the planets
through the analysis of transit time variation (TTV) signals
\citep{Agol2005MNRAS.359..567A,Holman2005Sci...307.1288H,Steffen2012PNAS..109.7982S}
with a photo-dynamical approach (Sec.~\ref{sec:analysis})
on a data-set almost two years longer than \citet{Korth2023A&A...675A.115K}.
We finally present the results in Sec.~\ref{sec:discussion} 
and in Sec.~\ref{sec:conclusions} we compare them with \citet{Korth2023A&A...675A.115K} work
and draw our conclusions.


\section{Observations}\label{sec:obs}

\subsection{Photometry}\label{sec:phot}

\begin{table*}
    \centering
    \caption{\label{tab:obslog}Log of TOI-1130 observations.}
    \small
    \include{obs_log}
\end{table*}

\subsubsection{CHEOPS}\label{sec:cheops}
TOI-1130 has been observed with CHEOPS within three GTO programs for a total of 17 light curves 
(see Table~\ref{tab:obslog} for the full list of the CHEOPS observations).
In particular, planet b has been observed in 11 visits, initially within the program
\textit{CHESS}\footnote{CHEOPS GTO PR-100031, M. Hooton}
and later in the program \textit{System Architecture}\footnote{CHEOPS GTO PR-120053, A. Leleu}.
We collected six CHEOPS visits of TOI-1130 c, that was part of the GTO program
\textit{Companion to Warm Jupiter planets TTV}\footnote{CHEOPS GTO PR-100015, G. Piotto \& L. Borsato}\citep{Borsato2021MNRAS.506.3810B} and
also of the \textit{System Architecture} program.
For each visit we extracted the aperture photometry with default aperture of $25$~pixels provided by the DRP 14.1 \citep{Hoyer2020A&A...635A..24H}.
After each observation we analysed as single-visit the data with \pycheops{} \citep{Maxted2022MNRAS.514...77M},
with priors on the transit parameters based on the discovery paper \citep{Huang2020ApJ...892L...7H} and
we determined the detrending parameters using the Bayes Factor, 
as described in \citet{Maxted2022MNRAS.514...77M}.
This allowed us to update our linear ephemeris improving our prediction of the transit times
and constraining the observations with CHEOPS.
However, due to the large transit time variation (TTV) of both planets,
in the visit of planet c of the 2022-06-21 (CHEOPS visit num. 14 in Tab.~\ref{tab:obslog})
the post-transit show an additional transit-like
feature that, after many attempts on correcting it, we attributed to planet b.\par

\subsubsection{TESS}\label{sec:tess}
TOI-1130 has been observed by the Transiting Exoplanet Survey Satellite \citep[TESS;][]{Ricker2015JATIS...1a4003R}
in three sectors: 13, 27, and 67 (see Table~\ref{tab:obslog}).
For sector 13 (S13), we used the already flattened \texttt{KSPSAP\_FLUX} light curve from the
Quick Look Pipeline \citep[QLP;][]{HuangQLP2020RNAAS...4..204H},
while for sectors 27 (S27) and 67 (S67), we used the 
Presearch Data Conditioned Simple Aperture Photometry (\texttt{PDCSAP}) light curves (in the 20~s fast mode),
as processed by the Science Processing Operation Center \citep[SPOC;][]{jenkins2016}.\par

\subsubsection{Ground-based photometry}\label{sec:groundphot}

The ground-based observations used in our analysis were taken by six ground-based telescopes, namely
ASTEP\footnote{Antarctic Search for Transiting ExoPlanets}, 
CDK14\footnote{El Sauce Observatory in Chile}, 
PEST\footnote{Perth Exoplanet Survey Telescope, backyard observatory in western Australia}, and 
LCO-CTIO\footnote{Cerro Tololo Interamerican Observatory in Chile}, 
LCO-SSO\footnote{South at Siding Spring Observatory in eastern Australia},
and 
LCO-SAAO\footnote{South African Astronomical Observatory}
from the Las Cumbras Observatory (LCO).
The observations we used are largely the same as those used by \citet{Korth2023A&A...675A.115K} with four differences that will be discussed here.
Firstly, we decided to include the observations of ASTEP (0.4\,m) and PEST (0.3\,m) in addition to the observation of LCO-SSO (1\,m) of the same transit of planet c on 5 August 2020. 
Secondly, we fit for transits of both planet b and planet c in the observation of LCO-CTIO on the 26th of June 2021, while \citet{Korth2023A&A...675A.115K} do not report to have used the transit of planet b from this observation. 
Thirdly, we did not take into account the observation of planet b from LCOGT-SAAO on 7 October 2021, as this transit was observed simultaneously by CHEOPS. Given that the CHEOPS photometric precision is far greater than that of LCOGT-SAAO, 
and the fact that all contact points of this transit were well constrained by CHEOPS' observation, 
the inclusion of this data would unnecessarily complicate the analysis.
Lastly, we did not use LCOGT-CTIO's observation of planet b on 8 October 2021,
as it contained more red noise than any other ground-based observation and 
was the only data set for which no cotrending basis vectors were provided at the time of analysis. 
In summary, we used 19 ground-based telescope observations, encompassing 6 transits of planet b and 10 transits of planet c. 
A more detailed description of the data selection and processing is given in \cite{Degen2022}.
\par

We also obtained photometric observation with ASTEP+ \citep{2015AN....336..638G}, taking advantage of the new camera system with simultaneous observations in two bands, ASTEP+B between 400 and 700nm, and ASTEP+R between 700 and 1000nm \citep[see][]{astep+2022SPIE12182E..2OS}. We observed
two transits of planet b and three of planet c (see observation log in Table~\ref{tab:obslog}),  
for a total of 10 light curves. The data analysis was performed using aperture photometry, as described in \cite{2016MNRAS.463...45M}. 
For each light curve we also have some diagnostics, i.e., $X-Y$ coordinates on the CCD,
FWHM, the sky background, and the airmass.\par

\subsection{Radial velocities}\label{sec:rv}

We collected all the radial velocities (RVs) available in literature,
21 RVs with CHIRON \citep{CHIRON2013PASP..125.1336T} from the discovery paper by \citet{Huang2020ApJ...892L...7H},
49 RVs from HARPS \citep{HARPS2003Msngr.114...20M} and
20 from PFS \citep{Crane2006SPIE.6269E..31C, Crane2008SPIE.7014E..79C, Crane2010SPIE.7735E..53C} published by \citet{Korth2023A&A...675A.115K}.

\section{Stellar parameters}\label{sec:star}

We utilised our co-added high-resolution HARPS spectra ($R=115\,000$)
to perform spectroscopic modelling of TOI-1130. 
We began by running the empirical SpecMatch-Emp \citep{2017ApJ...836...77Y}
software which compares a library of well-characterised stars to our observations. 
The results indicate that the star is a K$6\,$V star.
We compared the outcome with Spectroscopy Made Easy\footnote{\url{http://www.stsci.edu/~valenti/sme.html}}
\citep[SME;][]{vp96, pv2017} which fits observations to compute synthetic spectra
from stellar atmosphere grids and atomic and molecular line data from VALD \citep{Ryabchikova2015}.
For modelling of TOI-1130, we chose the MARCS model atmosphere \citep{Gustafsson08},
and verified the results with the Atlas12 atmosphere grid \citep{Kurucz2013}.
We fixed the micro- and macro-turbulent velocities, $V_{\rm mic}$ and $V_{\rm mac}$,
to 0.1~km~s$^{-1}$ and 1.0~km~s$^{-1}$ \citep{gray08}.
The modelling steps are described in \citet{2018A&A...618A..33P}.\par

Even though mid-K stars and later are often challenging to model with spectral synthesis software,
the results of the final SME model are in very good agreement with Specmatch-emp.
The results from both models are listed in Table~\ref{tab:spectroparams}.
We started from both these two sets of outcomes to derive the isochronal parameters of the star (see below)
and we found that the SME-based data enable us to compute a more precise stellar mass. 
Therefore, we assumed the SME-based spectroscopic parameters as the reference values of this work.\par

To determine the stellar radius of TOI-1130 we used an MCMC modified infrared flux method
\citep{Blackwell1977,Schanche2020}.
Using spectral energy distributions (SEDs) built from stellar atmospheric models \citep{Castelli2003}
with priors coming from our spectral analysis, 
we computed synthetic photometry that we compared to observed broadband photometry in the following band-passes:
\textit{Gaia} $G$, $G_\mathrm{BP}$, and $G_\mathrm{RP}$, 2MASS $J$, $H$, and $K$, and \textit{WISE} $W1$ and $W2$
\citep{Skrutskie2006,Wright2010,GaiaCollaboration2022}.
Using the derived stellar bolometric flux and known physical relations,
we determined the effective temperature and angular diameter of TOI-1130
that we translated into stellar radius using the offset-corrected \textit{Gaia} parallax \citep{Lindegren2021}
to $R_\star = 0.697 \pm 0.011\ R_\odot$.\par

We used $T_{\mathrm{eff}}$, [Fe/H], and $R_{\star}$
along with their uncertainties as the basic set of input parameters
to then derive the isochronal mass $M_{\star}$ and age $t_{\star}$
by employing two different sets of stellar evolutionary models. 
In detail, we computed a first pair of mass and age estimates via the CLES code 
\citep[Code Li\`{e}geois d'\'{E}volution Stellaire;][]{scuflaire2008}, 
which generates the best stellar evolutionary track 'on-the-fly' 
accounting for the input parameters and following the Levenberg-Marquadt minimisation scheme \citep{salmon2021}.
For the second pair of estimates, instead, we utilised the isochrone placement algorithm \citep{bonfanti2015,bonfanti2016},
which interpolates the input values within pre-computed grids of 
PARSEC\footnote{\textsl{PA}dova and T\textsl{R}ieste \textsl{S}tellar \textsl{E}volutionary \textsl{C}ode: \url{http://stev.oapd.inaf.it/cgi-bin/cmd}}
v1.2S \citep{marigo2017} isochrones and tracks.
As the isochrone placement implements also the gyrochronological relation by \citet{barnes2010}
to work in synergy with isochrone fitting \citep[see][]{bonfanti2016},
we further inputted $v\sin{i}$ to improve convergence.
We finally checked the mutual consistence of the two respective pairs of outcomes via the $\chi^2$-based criterion
outlined in \citet{bonfanti2021} and combined the results obtaining
$M_{\star}=0.722_{-0.037}^{+0.042}\,M_{\odot}$ and $t_{\star}=5.4_{-4.9}^{+5.7}$ Gyr,
and a stellar density of $\rho_\star = 2.13 \pm 0.16\, \rho_\sun$.

\input{stellar_parameters}

\section{Data analysis and modelling}\label{sec:analysis}

For each TESS sector we discarded data-points with QUALITY factor greater than 0
and with flux 10-$\sigma$ above the median flux.
We then selected a portion of the light curve around each transit of both planets,
taking into account the transit duration and at least the equivalent of three CHEOPS orbits (about $98.77$ minutes each).
We sought the transits through the linear ephemeris and transit parameter by \citet{Huang2020ApJ...892L...7H},
and we visually checked if the transit was missed due to the predicted transit timing variation (TTV)
and we adjusted the centre of the portion if needed.
In one case we had to join a planet b and c transit because the two portions were too close to keep them separated.
We repeated the same procedure and portioned S67 with updated linear ephemeris from \citet{Korth2023A&A...675A.115K} and CHEOPS data.
\par

As the results of the shallow transit of b and of the gaps in the CHEOPS visits,
during the preliminary single-visit analysis we found that the planetary parameters
did not agree visit-by-visit.
Given the high number\footnote{
When detrending 17 CHEOPS light curves, more than 90 parameters are required, 
with approximately 11 common parameters and 5 per visit.
Taking into account all the photometries, radial velocity datasets, stellar parameters, 
and orbital parameters we could have more than 250 parameters.
}
of parameters needed to perform an MCMC analysis of a
simultaneous fit of the transits and the detrending parameters for each light curve,
we decided to inspect which are the common diagnostics across all the light curves.
Plotting as function of the roll angle ($\phi$) the different diagnostics,
we found that the position on the CCD ($x-y$ offset), and its second derivative,
could be treated as common parameters across all the visits.
We also modelled the roll angle with a common sinusoidal with six harmonics.
While a flux constant, a linear and quadratic term in time, the background, 
and the ramp effect \citep[see][]{Maxted2022MNRAS.514...77M} 
were defined visit-by-visit.
We masked all the portions of the transits in the light curves, 
we performed a \pycheops{} detrending-like least-squares\footnote{
Levenberg-Marquardt algorithm of MINPACK \citep{MINPACK-1} implemented in
\href{https://docs.scipy.org/doc/scipy/reference/generated/scipy.optimize.leastsq.html\#scipy.optimize.leastsq}{scipy.optimize.leastsq}
}
fit of all the visits simultaneously.
We also extract the ``PSF imagette photometric extraction'' (\pipe) 
package\footnote{\url{https://github.com/alphapsa/PIPE}} photometry,
that is less prone to contamination and background effects
\citep{morris2021A&A...653A.173M,brandeker2022A&A...659L...4B}.
We apply the same full-detrending, after masking the transits.
We repeated this analysis with and without the last CHEOPS visit with planet c and b (hereafter V14).
We found that V14 cannot be detrendend in such way, so we kept the analysis without it.
We also found that the PIPE photometry provided a median out-of-transit standard deviation 
$\sigma_\mathrm{phot, PIPE} = 552 \pm 23$~ppm, 
lower than DPR case with $\sigma_\mathrm{phot, DRP} = 659 \pm 58$~ppm.
For this reason, in following analysis we decided to use 16 pre-detrended visits
from PIPE photometry out of 17, and the V14 from PIPE without pre-detrending.
\par

\subsection{Ground based}\label{sec:ground_analysis}

Ground-based observations generally exhibit higher levels of both white and red noise than space telescopes, mainly due to atmospheric turbulences. Reducing the effect of these systematic errors requires the introduction of detrending parameters, thereby increasing the dimension of the parameter space. The increase in dimensionality not only increases the computational complexity but also makes statistical inference challenging, e.g. by slowing convergence and introducing degeneracies, thus limiting the comprehensive exploration of the parameter space and, therefore, the reliable extraction of meaningful information. For these reasons, we decided to prioritise precision over quantity and split the analysis into two phases: first, a photometric analysis of the ground-based telescope data to extract the transit times of the planets, and then a photo-dynamic analysis of the space telescope data. In this way, we are able to incorporate the transit times obtained from the ground-based telescope data into the photo-dynamic analysis without inflating the dimension of the parameter space of the main analysis by more than 100 parameters that were identified as useful for detrending the ground-based telescope data by the Bayesian information criterion. In the following subsection we give a brief overview of the analysis of the ground-based telescope data, which is described in more detail in \cite{Degen2022}.

We performed two photometric analyses, each combining the ground-based observations with data from TESS sectors 13 and 27. The difference between the two analyses lay in the strictness with which we discarded flux measurement points as outliers, which was done on the basis on their absolute deviation from the local median with a width size of 11.
The two TESS sectors were included to constrain nuisance parameters, in particular the planetary radii, and, thus, reduce the uncertainty in the extracted transit times.
The analyses were performed using the python package \textsc{exoplanet} \citep{ForemanMackey2021, exoplanet:zenodo}
and its dependencies \citep{Agol2020,
Kumar2019, Collaboration2013, Collaboration2018a,
Luger2019, Salvatier2015, Team2016}. \textsc{PyMC3} \citep{Salvatier2015} served as the inference engine, employing the No U-Turn Sampler (NUTS) for efficient posterior exploration. Convergence was assessed using Gelman-Rubin diagnostics, trace plots, and corner plots. 

We restricted our analyses from small to moderate eccentricities ($< 0.2$), so we should remain unaffected from the divergent errors that the modified Newton-Raphson method used by this package has been shown to have at large eccentricities \citep{Tommasini2021}.
The eccentricity and argument of periapsis were parameterised using 
$\sqrt{e}_p \cos(\omega_p)$ and $\sqrt{e}_p \sin(\omega_p)$ for enhanced efficiency. 
Informative Gaussian priors were chosen for stellar mass and radius as well as the quadratic limb darkening coefficients of TESS. 
For the latter, we fitted a Gaussian to the coefficient estimates from Atlas and PHOENIX provided by \cite{claret2017limb}, 
such that all coefficients with 
$T_\mathrm{eff} \in [4000, 4500]$, 
[Fe/H] $\in [0, 0.2]$, 
$\log{g} \in [0, 0.2]$ lie in the $R(f_{5\,\%})$
region of highest density of the prior, 
defined as 
$R(f_\alpha) = \bigl\{x: \mathbb{P}[x] > f_\alpha \bigr\}$
such that $P[x \in R(f_\alpha)] = 1-\alpha$.
The quadratic limb darkening coefficients for the filters of the ground-based telescopes were sampled uniformly according to the sampling scheme of \cite{Kipping2013}.
Weakly informative Gaussian priors were chosen for the transit times.
Gaussian processes were used to model the stellar variability and instrumental trends in the TESS data. 
We also included a photometric jitter term for each ground-based observation and each TESS sector, 
which was added in quadrature to the respective flux uncertainties. 
Each of these jitter terms was sampled from a log-normal distribution, 
with a mean given by the standard deviation of the out-of-transit flux of the respective observation and a standard deviation of two.\par

The two different outlier removal strategies had little effect on the transit times inferred from the TESS data, but resulted in slight discrepancies between the posterior transit times of ground-based telescopes for three transits of planet b and one transit of planet c. In these cases, the posterior means of the two models were separated by more than one standard deviation of either distribution, resulting in differences up to 5.12 minutes for planet b and 1.3 minutes for planet c. We, thus, find that, at least in these cases, our posterior uncertainties underestimate the uncertainties that would arise if one were to marginalise across different strategies to remove outliers. Although marginalising over different outlier removal strategies would be desirable, it is computationally too expensive. Accordingly, we decided to use the mean and standard deviation of the stricter outlier removal strategy, i.e. the strategy in which more points were rejected as outliers, given that its the posterior transit time distributions had larger standard deviations in the discrepant cases.

\subsection{Transit modelling and transit timing variations}\label{sec:modeltra}

We used a two-step approach to determine the initial parameters for the photo-dynamical
analysis of the following Section~\ref{sec:photodyn}.\par

Firstly, we analysed, simultaneously,
pre-detrended CHEOPS photometry, 
portioned TESS light curves,
and new ASTEP+ photometry\footnote{
These transits have been observed after last TESS sector and after the analysis of the older
ground-based observations, so we decided to analyse them with the photo-dynamical approach.
}
with \pyorbit{} \citep{Malavolta2016A&A...588A.118M, Malavolta2018AJ....155..107M}.
We used stellar parameters from Section~\ref{sec:star} as Gaussian priors, 
and we computed quadratic limb-darkening (LD) with 
\textsc{PyLDTk} \citep{Husser2013, Parviainen2015} for CHEOPS and TESS,
while we decided to use Gaussian priors on quadratic LD coefficients 
$u_{1} = 0.37\pm0.1$ and $u_{2} = 0.25\pm0.1$ for ASTEP+, both $B$ and $R$ filters.
We assumed fixed periods from previous incremental analysis 
(that is periods used to schedule CHEOPS visits)
and circular orbits.
We included CHEOPS V14 with \pycheops-like detrending of
$x-y$ offset as $\rmd f/\rmd x$, $\rmd^{2}f/\rmd x^{2}$,
$\rmd f/\rmd y$, $\rmd^{2}f/\rmd y^{2}$, $\rmd^{2}f/\rmd x \rmd y$,
background ($\rmd f/\rmd bg$),
three harmonics of the satellite roll angle ($\phi$) as 
$\rmd f/\rmd \cos\phi$,  $\rmd f/\rmd \sin\phi$, 
$\rmd f/\rmd \cos2\phi$, $\rmd f/\rmd \sin2\phi$,
$\rmd f/\rmd \cos3\phi$, $\rmd f/\rmd \sin3\phi$,
a flux constant ($f_{14}$), and a linear term in time ($df/dt$).
We detrended each ASTEP+ light curves for each filter taking into account
a flux constant ($f$), a quadratic term ($\rmd f/\rmd t$, $\rmd^{2}f/\rmd t^{2}$),
the CCD position ($\rmd f/\rmd x$, $\rmd f/\rmd y$), 
FWHM value ($\rmd f/\rmd fwhm$), and sky background ($\rmd f/\rmd sky$).
The transits have been modelled with the \textsc{batman} package \citep{Kreidberg2015PASP..127.1161K},
and for the TESS Sector 13 we used a super-sampling factor of 30.
The planetary and transit parameters of each planet have been shared across all the light curves, filters, and telescopes.
We let free all the transit times ($T_{0}$s) for each transit light curve.\par
\begin{figure*}
\resizebox{\hsize}{!}{\includegraphics{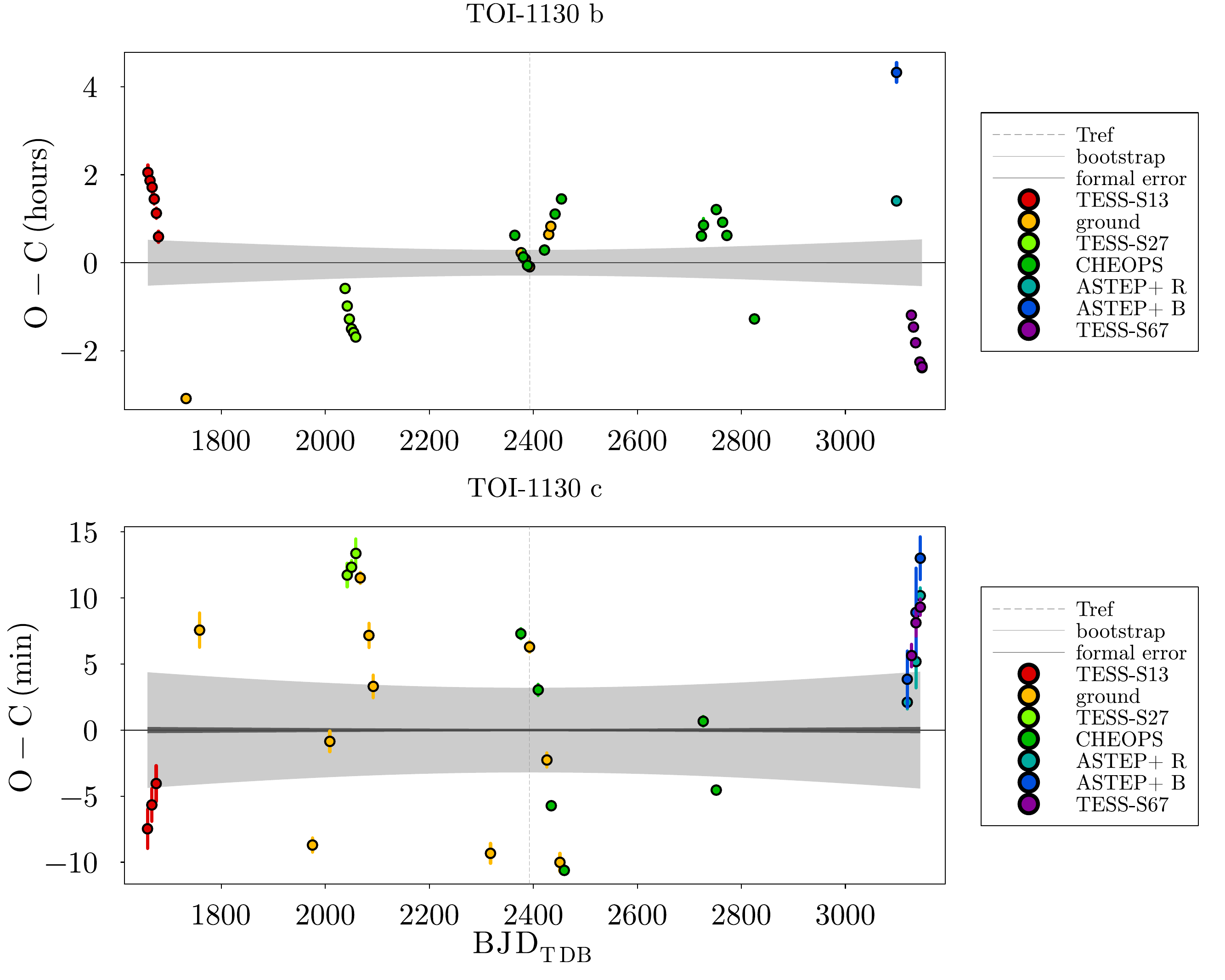}}
\caption{Observed minus Calculated (O-C) diagram of TOI-1130~b (top) and TOI-1130~c (bottom).
The Calculated $T_0$s were determined from the linear ephemeris:
$T_\mathrm{lin, b, N} = 2392.81470\pm0.00021\, (0.01260) + N \times 4.078668\pm0.000002\, (0.000097)\, \mathrm{BTJD_{TDB}}$,
and
$T_\mathrm{lin, c, N} = 2392.66995\pm0.00007\, (0.00218) + N \times 8.349560\pm0.000002\, (0.000023)\, \mathrm{BTJD_{TDB}}$,
with formal (thin-black area) and bootstrap (between round brackets, gray area) uncertainty.
Ground-based $T_0$s presented in Sec.~\ref{sec:groundphot} and analysed in Sec.~\ref{sec:ground_analysis}
are reported as ground.
$T_\mathrm{ref}$ is the reference time of the linear ephemeris plotted as dashed-gray vertical line.
All the times are in BTJD, that is $\mathrm{BJD_{TDB}}-2\,457\,000$.
}
\label{fig:oc}
\end{figure*}

We decided to run \pyorbit{} combining
the quasi-global optimiser \pyde{} \citep{StornP97, Parviainen2016}
with a population\footnote{
In a Differential-Evolution optimiser the population is the number of 
configurations set (or orbital configurations) for each iteration (or generation).
The population evolves with each iteration until the stopping criterion is reached.
} of 328 for 50\,000 generations 
and the \emcee{} package \citep{Foreman2013, DFM2019JOSS....4.1864F}
with 328 walkers for 2\,000\,000 steps.
We removed the first 1\,200\,000 steps as burn-in, 
after checking the convergence of the chains by
the Gelman-Rubin \citep{GelmanRubin1992} $\hat{R} < 1.01$
and the auto-correlation function \citep{DFM2019JOSS....4.1864F},
and we applied a thinning factor of 1\,000 to reduce CPU and memory load.
We take as best-fit parameters the maximum a posteriori (MAP) set,
and we extract the $T_{0}$s (see Table~\ref{tab:T0s_b} and \ref{tab:T0s_c})
of each light curve and planet.
We fitted new linear ephemeris 
(see Table~\ref{tab:T0s_b} and \ref{tab:T0s_c} for planet b and c, respectively) and
we created the observed minus calculated (O-C) diagrams 
(see Fig~\ref{fig:oc})
to assess the TTV signals.\par

\begin{table}
    \centering
    \caption{\label{tab:T0s_b}Transit times ($T_0$s) of TOI-1130 b from the photometric analysis with \pyorbit{}
    and
    as described in Sec.~\ref{sec:ground_analysis}.
    }
    \include{T0s_b}
\end{table}

\begin{table}
    \centering
    \caption{\label{tab:T0s_c}Transit times ($T_0$s) of TOI-1130 c as in Table~\ref{tab:T0s_b}.}
    \include{T0s_c}
\end{table}

Then, we run a dynamical analysis with \trades\footnote{\url{https://github.com/lucaborsato/trades}}
\citep{Borsato2014A&A...571A..38B, Borsato2019MNRAS.484.3233B, Borsato2021MNRAS.506.3810B}
using as observable the $T_{0}$s and the boundaries for some parameters from the \pyorbit{} analysis.
We let run \trades{} with \pyde{} with a population of 96 parameter set for 90\ 000 generations.
Differently from previous analysis, 
we let vary also the eccentricities, $e$ (and associated argument of pericenter, $\omega$),
and held fix the radii and the inclination of the planets.
Fitting parameters (e.g. $\sqrt{e}\cos\omega$ and  $\sqrt{e}\sin\omega$), boundaries, and priors 
were defined as in \citet{Nascimbeni2023A&A...673A..42N}.
At the end of the analysis we extracted the best-fit solution and
computed the masses and physical parameters,
that we used, jointly with \pyorbit{} results, as initial parameters for the successive
photo-dynamical approach.\par

\subsection{Photo-dynamical modelling}\label{sec:photodyn}

The photo-dynamical approach allows us to simultaneously fit the transit photometry (with detrending),
transit times, and radial velocities, during the dynamical integration of the full system.
An upgraded version of \trades{} (\photrades{}) models transits with the \pytransit{} 
package\footnote{\url{https://github.com/hpparvi/PyTransit}}
\citep{Parviainen2015MNRAS.450.3233P}.
The code integrates the orbits of the planets and computes all the possible transit times
with associated Keplerian elements.
Then, it automatically (and blindly) select the transit times and orbital elements of all the planets
for each photometric portions, allowing us to model more planets for the same light curve portions,
and pass them to \pytransit{}.
It does not take into account for planet-planet (mutual) occultation.
However, to reduce the computational time and the number of parameters,
we decided to use the photometry (with detrending when needed)
as in the \pyorbit{} analysis (see Sec.~\ref{sec:modeltra}),
and only the $T_{0}$s for the transit from ground-based facilities
analysed in Sec.~\ref{sec:ground_analysis}.
For each radial velocity data-set we included an
offset ($\gamma$) and jitter term (in $\log_{2}$),
and a common linear trend in time \citep[as in][]{Korth2023A&A...675A.115K}
to take into account the possible influence of an additional planet
on far outer orbit.
See all 116 fitted parameters with boundaries and priors in Table~\ref{tab:phototrades}.\par
\begin{figure*}
\resizebox{\hsize}{!}{\includegraphics{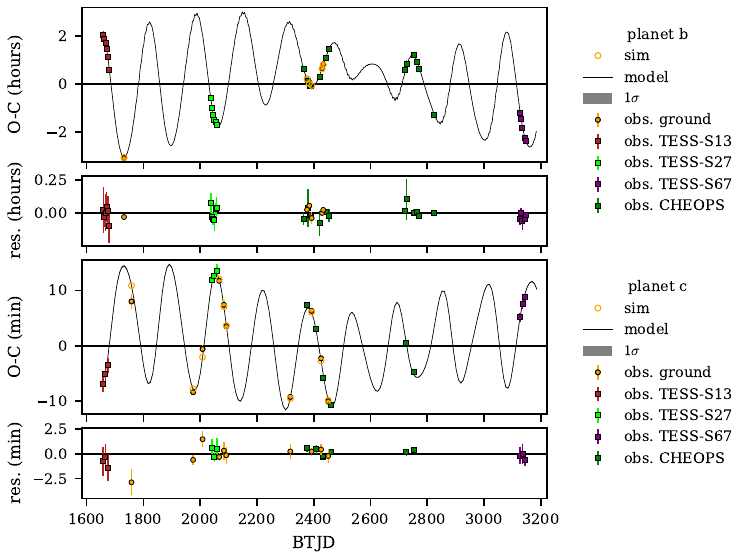}}
\caption{Observed minus Calculated ($O-C$) diagram of TOI-1130~b (top) and TOI-1130~c (bottom) from
the photo-dynamical fit with \trades{} of the ground-based $T_{0}$s (orange-black circles, simulated ones as open-orange circles) of
photometry 
described in Sec.~\ref{sec:ground_analysis}.
The $O-C$ has been computed with the linear ephemeris in Table~\ref{tab:T0s_b} and \ref{tab:T0s_c}.
Observed transit times of TESS and CHEOPS, not used in the fit,
have been over-plotted (as colour-coded squares) on the $O-C$ and residual panels,}
full MAP model as black line with $1\sigma$ gray-shaded area, that is almost invisible
because the model is very well constrained.
Times in $\mathrm{BTJD} = \mathrm{BJD_{TDB}}-2\,457\,000$.
\label{fig:oc_photodyn}
\end{figure*}
\begin{figure*}
\resizebox{\hsize}{!}{\includegraphics{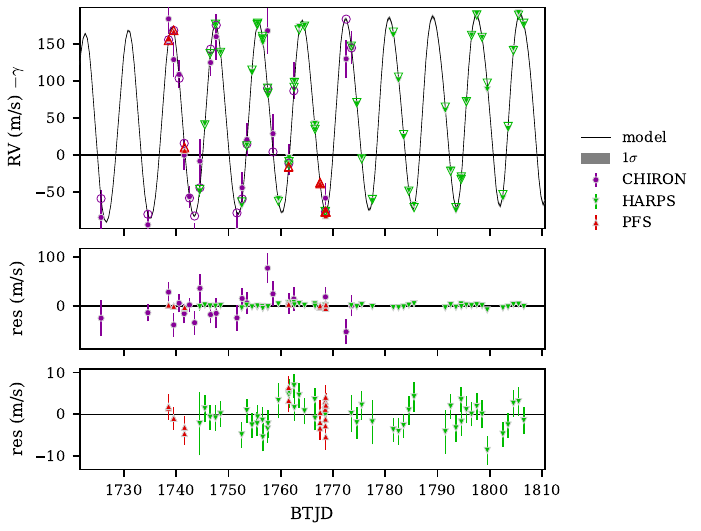}}
\caption{Radial velocity (subtracted RV offset, $\gamma$) plot in \textit{top panel}
with different marker and colour for each data-set.
Observations as filled-markers and corresponding simulations as open-markers;
full MAP model as black line with $1\sigma$ gray-shaded area.
Residuals in \textit{mid-panel} with all data-sets.
Residuals without the CHIRON dataset in \textit{bottom-panel}.
Times in $\mathrm{BTJD} = \mathrm{BJD_{TDB}}-2\,457\,000$.
}
\label{fig:rv_photodyn}
\end{figure*}
\begin{figure*}
\centering
\includegraphics[width=17cm]{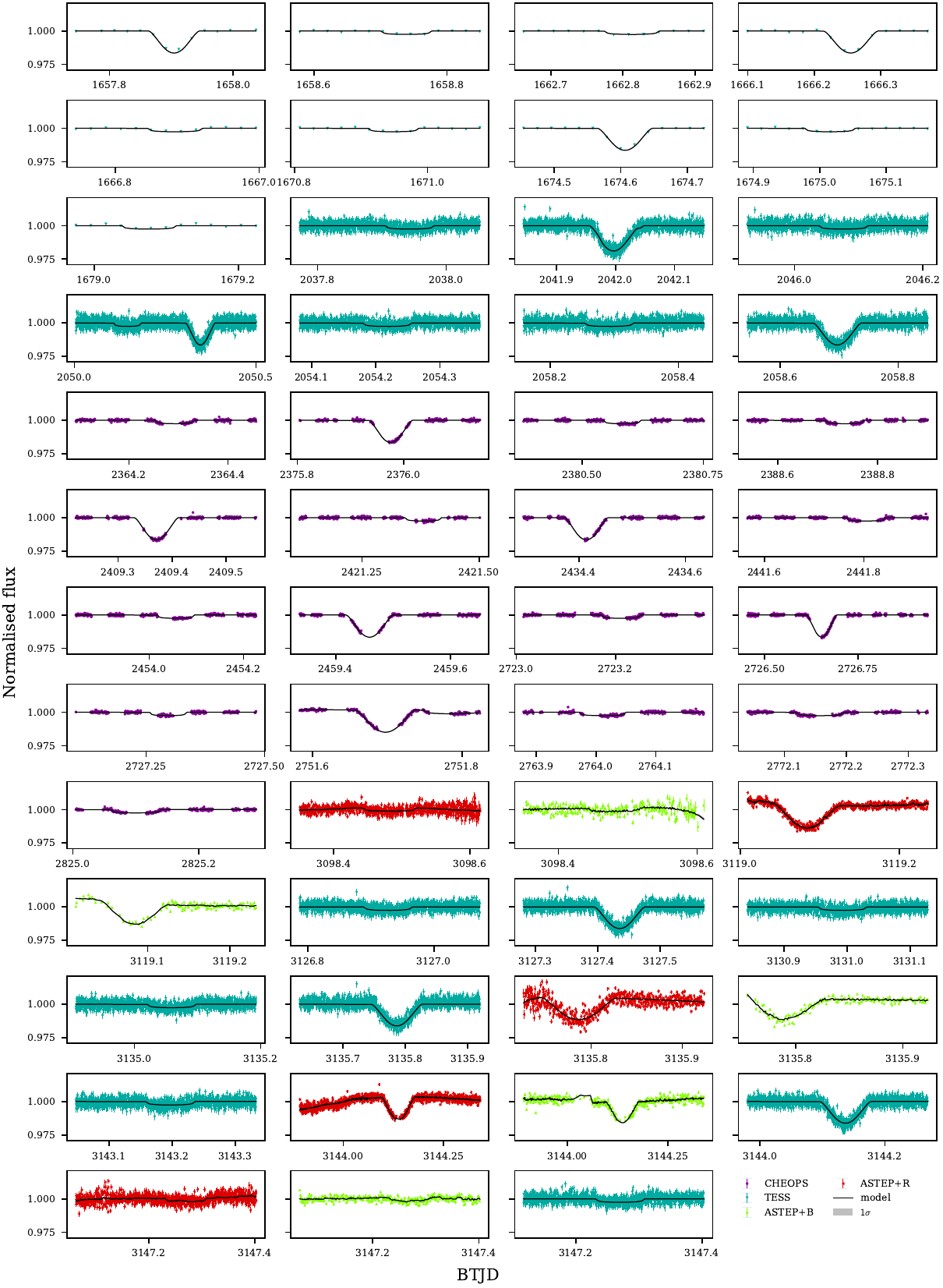}
\caption{Photometry of each light curve used in the photo-dynamical analysis.
Each panel has the data (filled markers, different marker and colour based on the facilities,
see legend) with error-bars,
and the over-plotted (over-sampled) full model (black line, transit and detrending) computed
with the MAP parameter set from \trades{}.
Model uncertainties plotted as gray-shaded area at $1\sigma$, but barely visible. 
Times in $\mathrm{BTJD} = \mathrm{BJD_{TDB}}-2\,457\,000$.
}
\label{fig:photo_photodyn}
\end{figure*}
\begin{figure*}
\centering
\includegraphics[width=17cm]{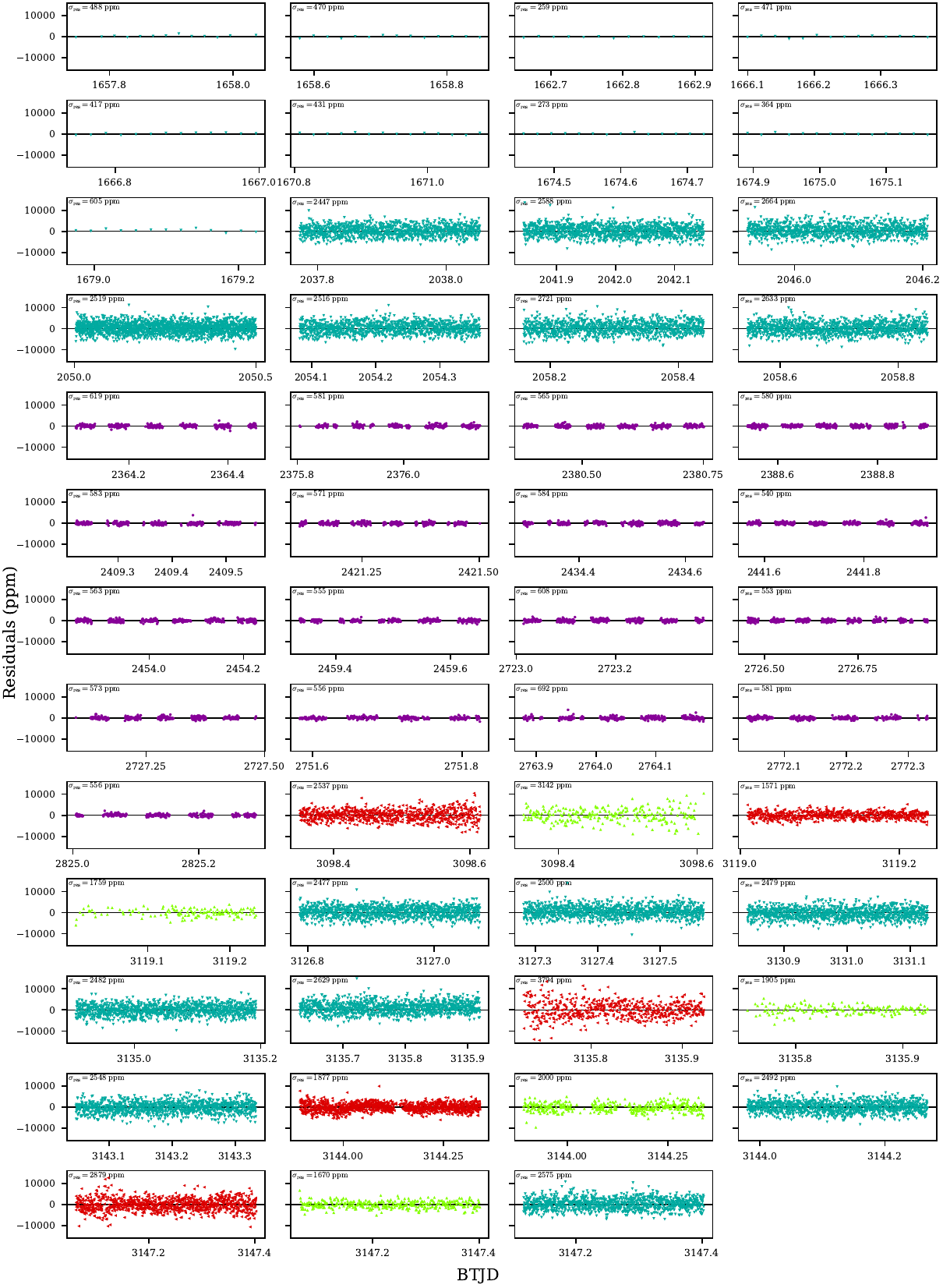}
\caption{As in Fig.~\ref{fig:photo_photodyn}, but for the residuals (in ppm) between
the observed photometry and the MAP full model.
Each panel shows in the upper-left corner the standard deviation of the residuals in ppm.
}
\label{fig:res_photodyn}
\end{figure*}

We run \photrades{} with \emcee{} with 232 walkers for $2\ 000\ 000$ steps
with a thinning factor of 100.
We discard the first $696\ 000$ steps as burn-in 
after checking the convergence of the chains through visual inspection and 
with Gelman-Rubin ($\hat{R}$),
Geweke \citep{geweke1991},
and ACF statistics.
As representative of the best-fit solution we took the
maximum a posteriori (MAP), 
the parameter set that maximise the log-probability .
Due to the high complexity of the problem, 
some MAP parameters (fitted or physical) 
are outside of the $1\sigma$ uncertainty
defined as the high density interval (HDI) at the 68.27\%.
In this case, we computed and reported the HDI at the 95.44\% as $2\sigma$ equivalent.
See the full report of the fitted and physical parameters in Table~\ref{tab:phototrades}.
See $O-C$ diagram of the ground-based $T_{0}$s in Fig.~\ref{fig:oc_photodyn},
RV plot in Fig.~\ref{fig:rv_photodyn},
and photometry in Fig.~\ref{fig:photo_photodyn}
(with photometric residuals in Fig.~\ref{fig:res_photodyn}).\par

We searched for additional signals in the Doppler data computing the 
generalised Lomb-Scargle\footnote{Implemented in \textsc{Julia} version available at \url{http://juliaastro.org/LombScargle.jl/stable/}.} periodogram \citep[GLS,][]{Zechmeister2009A&A...496..577Z} of the HARPS and PFS RV residuals. We found no significant signals in the power spectrum, with the highest peak in the GLS periodogram having a relatively high false alarm probability (FAP) of $\sim$10\%, as derived using a bootstrap randomisation approach with 10\,000 repetitions \citep{Hatzes2019}.

\include{phototrades_table}

\section{Discussions}\label{sec:discussion}

\subsection{Mean-Motion Resonance analysis}\label{sec:mmr}

\begin{figure}[!ht]
    \begin{center}
    \includegraphics[width=0.99\linewidth]{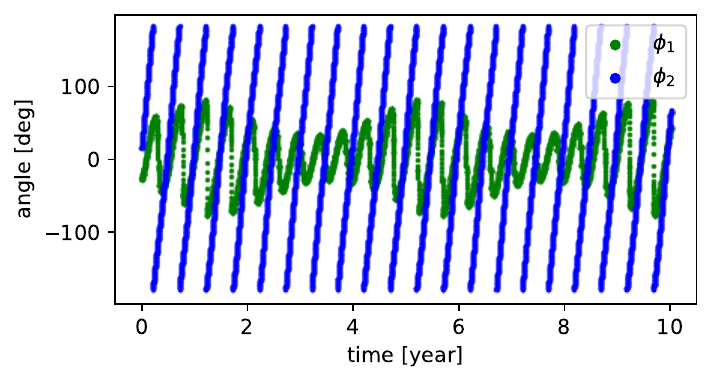}\\
    \includegraphics[width=0.99\linewidth]{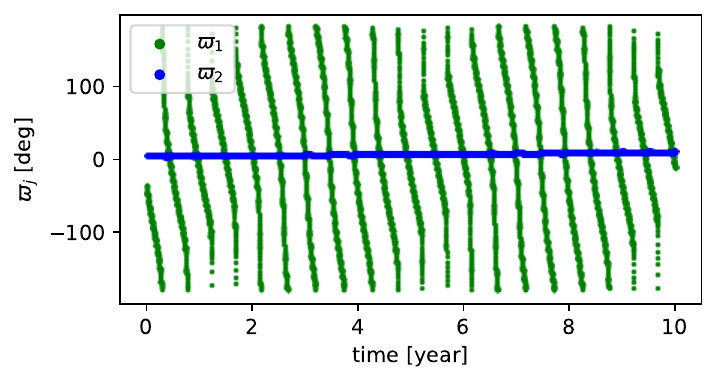}
    \caption{\label{fig:angles} 
    Example of evolution of the angles for a sample of the posterior.
    The top panel shows the resonant angles 
    $\phi_1 = \lambda_1 -2 \lambda_2 +\varpi_1$ and 
    $\phi_2 = \lambda_1 -2 \lambda_2 +\varpi_2$, 
    where 1 refers to planet b and 2 to planet c,
    $\lambda$ is the mean longitude and
    $\varpi$ the longitude of periastron of the planets.
    The bottom panel shows the evolution of the longitude of periastron.
    }
    \end{center}
\end{figure}

\begin{figure}[!ht]
\begin{center}
\includegraphics[width=0.99\linewidth]{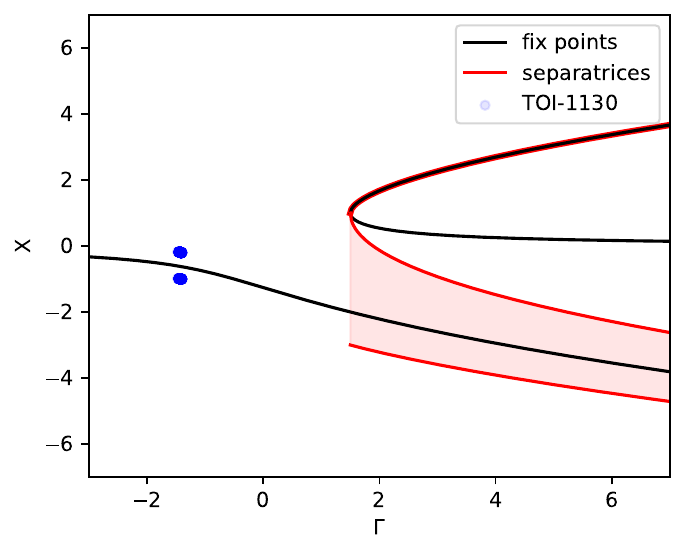}
\caption{\label{fig:phasespace} 1-degree of freedom model of the 2:1 MMR \citep{HenrardLemaitre1983,Deck2013}. $X$ and $\Gamma$ are function of the orbital elements and masses of the system. $\Gamma$ represent how deep the system is in the resonance, while $X$ parameterise the position of the fix points and separatrices. The blue dots represent the intersection of 300 randomly-selected samples of the posterior with the $X$-$\Gamma$ plane.  
}
\end{center}
\end{figure}

For a pair of planets close enough to a first order mean-motion resonance (MMR),
the expected TTVs is made of two terms: the evolution of the resonant angle ($\phi$),
whose period is roughly proportional to $(m/m_\star)^{-2/3}P_\mathrm{orb}$,
where $m$ is comparable to the mass of the planets,
and a slower term  linked to the evolution of the eccentricities and longitude of periastron \citep{NeVo2016}.
The relation between these two periods depend on the position of the system with respect to the exact resonance. Often, and in particular for smaller planets, an observation baseline of a few years only allow to observe the effect of the resonant angle, resulting in a mostly sinusoidal TTV signal. However, in the case of TOI-1130, the period of the two terms appear to be very close, with the period of precession of the inner orbit only slightly slower than the evolution of the resonant angle $\phi_1$, see Fig. \ref{fig:angles}. Adding up these two terms with similar period results in a TTV signal that appears to be modulated in amplitude, see Fig. \ref{fig:prediction}. 

The good constraints that we get for the resonant term of the TTVs allows us to constrain the resonant part of the architecture, as can be seen in Fig. \ref{fig:phasespace}. In that figure, we can see that the system lies outside the formal resonant domain (red area in the figure), unlike, for example, TOI-216 \citep{Nesvorny2022}. 
Observing the resonant term on its own does not allow to constrain the non-resonant part of the eccentricities, often resulting in a highly degenerated posterior for the eccentricities and longitudes of periastron \citep{Leleu+2021b,Leleu+2022,Leleu+2023}. However, for TOI-1130, we can see the effect of the evolution of the periastron and the eccentricities on the TTVs, resulting in good constraints on these parameters.

Assuming that the pair was initially captured into the 2:1 due to a convergent migration in the proto-planetary disc \citep[e.g.][]{Weidenschilling1985,TePa2007}, the observed architecture of the system enables in-depth study of its long term tidal evolution. Indeed, 
for planets whose orbital period is typically below 10 days, tides are expected to be an efficient mechanism to damp the eccentricity of the planet, effectively pushing the planets outside of the exact resonance \citep[e.g.][]{Novak2003,Lee2013,Delisle2014}.
\par

\subsection{Internal Structure Modelling}\label{sec:internal_model}

We modelled the interior structure of TOI-1130 b using a Bayesian inference scheme, following the method introduced in \cite{Leleu+2021}, which is based on \cite{Dorn+2017}. The forward model used is an early version of \cite{Haldemann+2024} and uses equations of state from \cite{Hakim+2018}, \cite{Sotin+2007} and \cite{Haldemann+2020} to model each planet as a spherically symmetric structure made up of an inner iron core with up to 19\% of sulphur, a silicate mantle made up of oxidised Si, Mg and Fe and a condensed water layer. On top of each such structure, a H/He envelope is modelled separately following \cite{Lopez+Fortney2014}. The elemental Si/Mg/Fe ratios are assumed to be stellar, following \cite{Thiabaud+2015}. While other studies find that at least for rocky planets this correlation might not be 1:1 \citep{Adibekyan2021Sci...374..330A}, the low density of TOI-1130 b, which means that it likely hosts a thick atmosphere, renders this assumption more reliable.

For the Bayesian inference, we use a uniform prior for the mass fractions of the inner core, mantle and water layer (on the simplex on which they add up to 1), with an upper limit of 0.5 for the water mass fraction \citep{Thiabaud+2014, Marboeuf+2014}. The mass of the H/He layer is sampled from a log-uniform prior. As the problem of modelling the internal structure of a planet is highly degenerate, the results of our analysis do depend to a certain extent on the chosen priors.

\begin{figure}
    \centering
    \includegraphics[width=\linewidth]{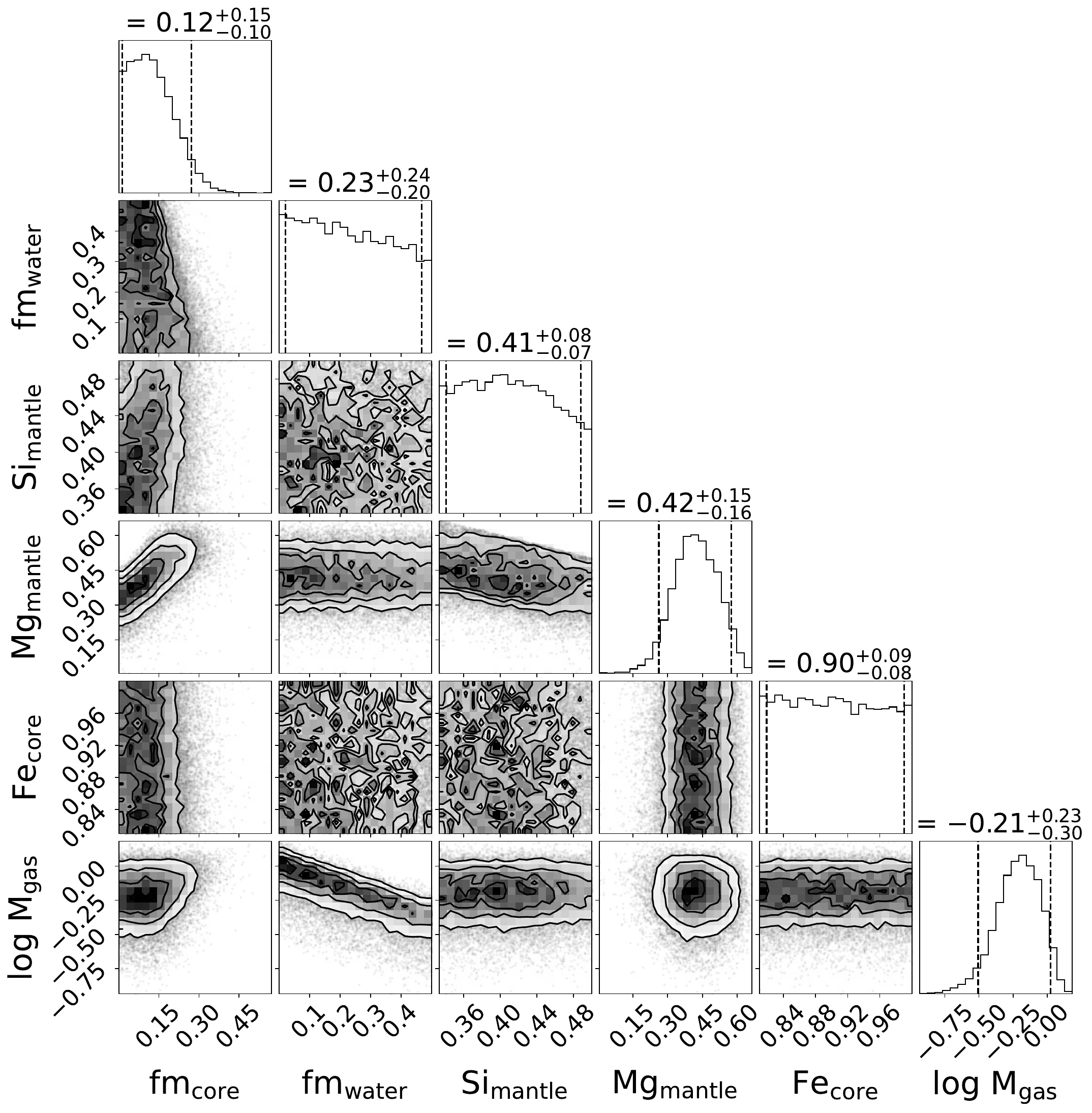}
    \caption{Inferred interior structure parameters of TOI-1130 b, with the values and errors above each histogram corresponding to the median and 5th and 95th percentile of each posterior distribution. The shown parameters are the mass fractions of the inner core and water layer (with respect to the solid part of the planet, they add up to 1 together with the mantle mass fraction), the elemental molar fractions of Si and Mg in the mantle and Fe in the inner core and the logarithm of the total H/He mass in Earth masses.}
    \label{fig:int_struct}
\end{figure}

The resulting constraints on the internal structure of TOI-1130 b are summarised in Figure \ref{fig:int_struct}. According to our model, the planet hosts a H/He envelope with a mass of $M_{\mathrm{gas}}=0.62^{+0.43}_{-0.31}$ $M_{\oplus}$ and a thickness of $R_{\mathrm{gas}}=1.26_{-0.25}^{+0.27}\,R_{\oplus}$, where the values and errors correspond to the median and 5th and 95th percentiles of the posterior distributions. 
The presence and mass of a potential water layer remains fully unconstrained. However, these results are affected by our model only considering water in condensed form and modelling the H/He envelope independently from the rest of the planet, thereby neglecting any pressure and temperature effects it has on the rest of the planet.

\section{Conclusions}\label{sec:conclusions}

In this work we improved our knowledge on the multi-planet system
TOI-1130 by re-analysing the whole transits and radial velocities presented
in \citet{Korth2023A&A...675A.115K} and with an additional TESS sector, six ground-based ASTEP+ transits,
and 17 new high-precision photometric data observed with the CHEOPS satellite.

From the full photo-dynamical model with \trades{}
we obtained stellar parameters, $R_\star$, $\rho_\star$, and $M_\star$, within $1\sigma$ of our priors,
but only $R_\star$ is consistent with the value derived by \citet{Korth2023A&A...675A.115K}.
This could be due to 
(1) different priors used (parameters determined with different methods and codes),
(2) different parameterisation (we fitted $R_\star$ and $\rho_\star$ instead of $R_\star$ and $M_\star$),
(3) and longer baseline ($\sim 2$~yr) of additional photometric data with very-high precision CHEOPS light curves.
\par

\begin{figure*}
\resizebox{\hsize}{!}{\includegraphics{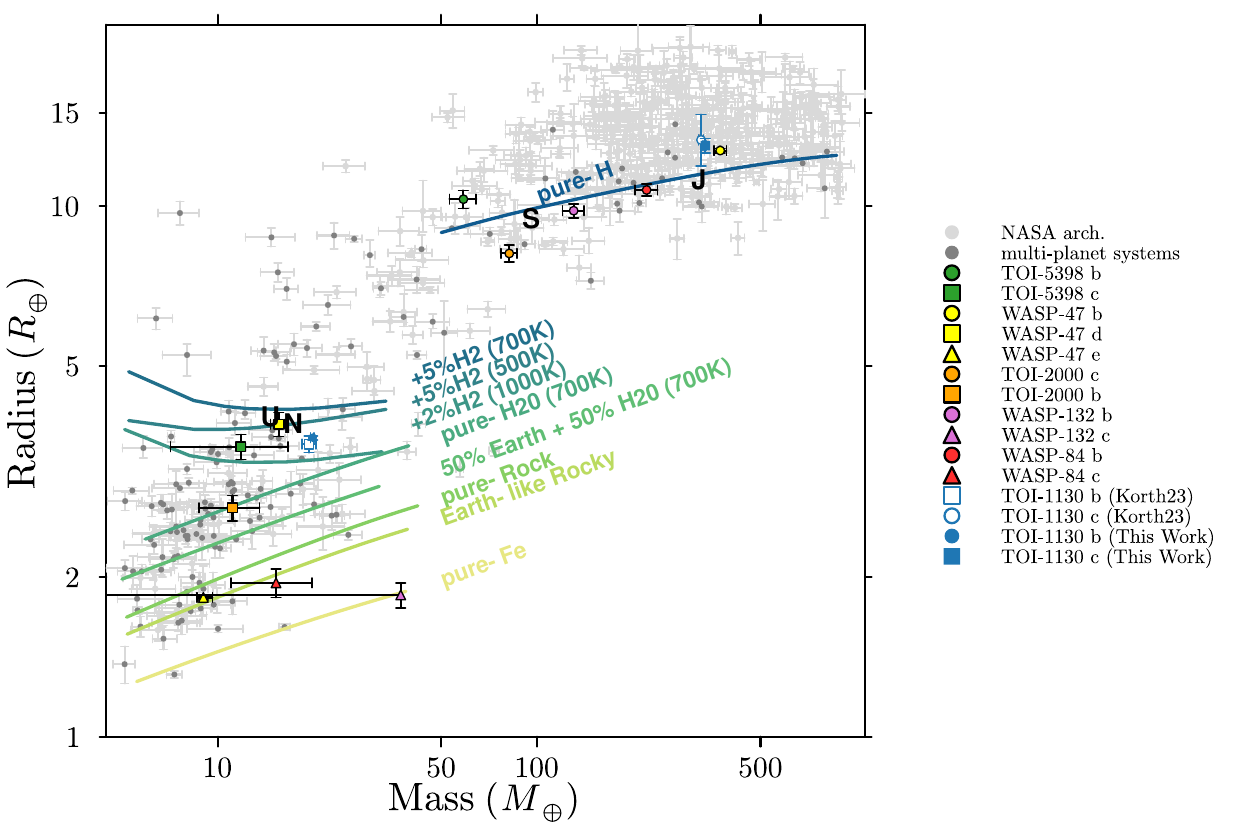}}
\caption{
    Mass-Radius relation plot with data from NASA Exoplanet archive
    for exoplanets with mass and radius precision lower than $20\%$.
    Models (plotted as color-coded lines) from
    \citet{Zeng2019PNAS..116.9723Z} and
    \url{https://lweb.cfa.harvard.edu/~lzeng/planetmodels.html\#mrrelation}.
    We plotted TOI-1130\,b as square, 
    white-fill blue-stroke from \citet{Korth2023A&A...675A.115K} and full-blue from this work;
    TOI-1130\,c as circle, same colour-code of TOI-1130\,b.
    Over-plotted 
    as darkgray circles the multi-planet systems and a sample of
    TOI-1130-like systems with different
    colours (based on host-name and with black outer stroke)
    and markers (based on planet type, 
    circle for giants, 
    triangle for $R_\mathrm{p} \leq 2\, R_\oplus$,
    and square in between):
    TOI-5398 \citep{Mantovan2023arXiv231016888M},
    WASP-47 \citep{Nascimbeni2023A&A...673A..42N},
    TOI-2000 \citep{Sha2023MNRAS.524.1113S},
    WASP-132 \citep{Hellier2017MNRAS.465.3693H,Hord2022AJ....164...13H},
    and WASP-84 \citep{Maciejewski2023MNRAS.525L..43M}.
    Over-plotted, also, Neptune (N), Uranus (U), Saturn (S), and Jupiter (J).
}
\label{fig:mr}
\end{figure*}

We found that $M_\rmb = 19.8_{-0.3}^{+0.2}\, M_\oplus$,
both radii ($R_\rmb = 3.66_{-0.04}^{+0.03}\, R_\oplus$, $R_\rmc = 13.0_{-0.4}^{+0.4}\, R_\oplus$)
and densities ($\rho_\rmb = 0.41_{-0.01}^{+0.01}\, \rho_\oplus$, $\rho_\rmc = 0.15_{-0.02}^{+0.01}\, \rho_\oplus$) 
agree with \citet{Korth2023A&A...675A.115K} within $1\sigma$,
while $M_\rmc = 336_{-5}^{+2}\, M_\oplus$ is consistent only at $2\sigma$.
Thanks to CHEOPS light curves with very high photometric precision,
and with additional ASTEP+ data,
the determined precision of each parameter is higher than \citet{Korth2023A&A...675A.115K},
in particular $\sigma_{M_\rmb} \lesssim 1.5\%$, $\sigma_{R_\rmb} \lesssim 1.1\%$,
$\sigma_{M_\rmc} \lesssim 1.5\%$, $\sigma_{R_\rmc} \lesssim 3\%$.
Those values imply a precision of $\sim 3\%$ and $\sim 11\%$
on the densities of planet b and c, respectively.
A comparison with \citet{Korth2023A&A...675A.115K} values and other similar systems
can be seen in the mass-radius relation in Fig.~\ref{fig:mr}.
We fitted eccentricities with flat-uniform priors in 
$(\sqrt{e}\cos\omega$, $\sqrt{e}\sin\omega)$ form and
we also found that both planets have slightly eccentric orbits
as suggested by \citet{Korth2023A&A...675A.115K},
but $e_\rmb$ is consistent within $2\sigma$, 
while we found a lower $e_\rmc$ consistent with their
value only at $4\sigma$.\par

Although we were unable to fit for an additional outer planet due to the current data, 
we were able to tentatively estimate the minimum mass ($M_\mathrm{d}\sin i$) and 
semi-major axis ($a_\mathrm{d}$) of candidate planet d
from the linear trend in the radial velocities.
The period $P_\mathrm{d}$ was assumed to be 161.762~days, 
which is twice the total time spanned by the RV observations.
Asymmetric Gaussian was generated for the stellar mass ($M_{\star}$) and 
RV linear trend from the photo-dynamical posterior.
The expected linear RVs were computed for each generated RV linear trend, 
and the semi-amplitude $K_\mathrm{RV}$ was estimated as the difference between 
the maximum and minimum values.
Inverting the equation 
$K_\mathrm{RV} \approx (M_\mathrm{d}\sin i)\ n\ a / M_\star$,
with the mean motion $n = 2 \pi / P_\mathrm{d}$,
and $a^{3} = G M_\star / n^{2}$,
we found 
$a_\mathrm{d} = 0.527 \pm 0.002$~au
and
$M_\mathrm{d}\sin i = 0.866_{-0.032}^{+0.025}\, M_\mathrm{Jup}$.
However, these values strongly depend on the assumption that the period of this candidate is 
approximately twice the time elapsed during RV observations. 
Observing only a linear trend in the RV,
it is possible that the period is longer, 
resulting in a higher mass.
The GLS analysis of RV residuals of PFS and HARPS does not exhibit a significant signal (FAP$\sim 10\%$). 
Further radial velocity measurements would increase the temporal baseline needed 
to detect the correct period of the outer candidate.
\par

\begin{figure}
\resizebox{\hsize}{!}{\includegraphics{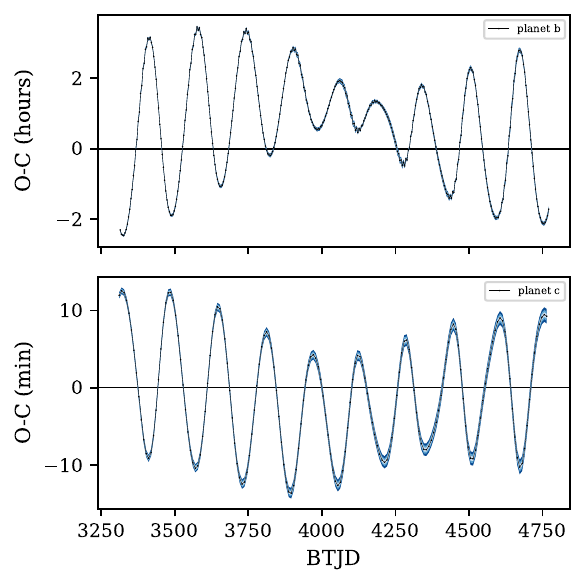}}
    \caption{O-C diagram of the synthetic $T_{0}$s predicted with the MAP orbital solution (black dots connected by black line)
    from photo-\trades{} analysis.
    Uncertainty plotted as $3\sigma$ in dark-blue, $2\sigma$ in light-blue, and $1\sigma$ whitish-blue.
    The $x$-axis in $\mathrm{BTJD} = \mathrm{BJD_{TDB}}-2\,457\,000$ ranges from 2024 until 2028.
}
\label{fig:prediction}
\end{figure}

The lower relative errors we obtained in the orbital parameters
allows us to predict the future transit times 
(see the O-C diagram in Fig.~\ref{fig:prediction}),
until 2028\footnote{We will provide the predicted transits with uncertainties as electronic form through CDS.},
with a very high precision ($\sigma_{T_{0},\mathrm{max}} = 2.6$~min and $20$~s for planet b and c, respectively).
This is very important because the precise knowledge of when both
planets transit is of fundamental importance for the upcoming
transmission spectroscopy observations 
with JWST \citep{Gardner_JWST_2023PASP..135f8001G} and Ariel \citep{Tinetti2018}.
In fact, TOI-1130 c is one of the planets with the highest 
Transmission Spectroscopic Metric \citep[TSM,][]{Kempton2018PASP..130k4401K},
and both planets are part of a JWST proposal\footnote{
JWST Proposal GO 3385, P.I. Chelsea Huang \citep{2023jwst.prop.3385H}.
}.
JWST and Ariel observations will allow us to characterise atmospheric abundances which are,
as suggested by \citet{Korth2023A&A...675A.115K},
crucial to understand the formation process that the system underwent. 
\par

\begin{acknowledgements}
    \input{acknowledgements.tex}
    ACMC ackowledges support from the FCT, Portugal, 
    through the CFisUC projects UIDB/04564/2020 and UIDP/04564/2020, 
    with DOI identifiers 10.54499/UIDB/04564/2020 and 
    10.54499/UIDP/04564/2020, respectively.\\
    \textit{Software.} The following software and packages have been used in this work: \textsc{numpy} \citep{numpy}, \textsc{matplotlib} \citep{matplotlib},  \textsc{scipy} \citep{scipy}, \textsc{h5py}, and
    \textsc{Julia}\footnote{\url{https://julialang.org/}} \citep{bezanson2017julia}.
\end{acknowledgements}

\section*{Data Availability}
Data will be available at CDS.
Data type: Observed-Fitted transit times in ascii file. 
ASTEP+ and CHEOPS-PIPE data and best-fit model in ascii file.
Predicted transit times from 2024 to 2028 in ascii file.

\bibliographystyle{aa} 
\bibliography{biblio}


\end{document}

%% file: obs_log.tex
\begin{tabular}{lllcrcccc}
    \hline\hline
    \noalign{\smallskip}
    \# & & DATA ID  & Start date & Duration & Frames & exp. time & Efficiency & Planet(s)\\  
       & &          & (UTC)      & (h)      &        & (s)       & (\%)       &          \\  
    \noalign{\smallskip}
    \hline
    \noalign{\smallskip} 
     - & TESS   & s0013-0000000254113311              & 2019-06-19T09:55 &    - &    - & 1800 &  - & b,c\\
     - & TESS   & s0027-0000000254113311-0189-a\_fast & 2020-07-05T18:31 &    - &    - &   20 &  - & b,c\\
     1 & CHEOPS & CH\_PR100031\_TG040601\_V0200       & 2021-05-29T14:06 &  8.7 &  260 &   60 & 50 & b\\
     2 & CHEOPS & CH\_PR100015\_TG018001\_V0200       & 2021-06-10T07:11 &  8.1 &  243 &   60 & 50 & c\\
     3 & CHEOPS & CH\_PR100031\_TG040801\_V0200       & 2021-06-14T20:58 &  8.9 &  309 &   60 & 58 & b\\
     4 & CHEOPS & CH\_PR100031\_TG040802\_V0200       & 2021-06-23T00:44 &  8.6 &  291 &   60 & 56 & b\\
     5 & CHEOPS & CH\_PR100015\_TG018101\_V0200       & 2021-07-13T17:09 &  8.0 &  237 &   60 & 49 & c\\
     6 & CHEOPS & CH\_PR100031\_TG042201\_V0200       & 2021-07-25T14:39 &  9.2 &  271 &   60 & 49 & b\\
     7 & CHEOPS & CH\_PR100015\_TG018201\_V0200       & 2021-08-07T18:59 &  8.1 &  246 &   60 & 51 & c\\
     8 & CHEOPS & CH\_PR100031\_TG044601\_V0200       & 2021-08-15T01:27 &  8.7 &  291 &   60 & 56 & b\\
     9 & CHEOPS & CH\_PR100031\_TG044602\_V0200       & 2021-08-27T08:08 &  9.2 &  308 &   60 & 56 & b\\
    10 & CHEOPS & CH\_PR100015\_TG018301\_V0200       & 2021-09-01T20:00 &  7.5 &  239 &   60 & 53 & c\\
    11 & CHEOPS & CH\_PR100031\_TG052801\_V0200       & 2022-05-23T12:13 &  8.7 &  275 &   60 & 53 & b\\
    12 & CHEOPS & CH\_PR100015\_TG022801\_V0200       & 2022-05-26T22:46 & 11.6 &  344 &   60 & 50 & c\\
    13 & CHEOPS & CH\_PR100031\_TG052901\_V0200       & 2022-05-27T14:17 &  9.1 &  247 &   60 & 45 & b\\
    14 & CHEOPS & CH\_PR120053\_TG004701\_V0200       & 2022-06-21T01:51 &  5.8 &  209 &   60 & 60 & b,c\\
    15 & CHEOPS & CH\_PR120053\_TG004601\_V0200       & 2022-07-03T08:57 &  7.2 &  240 &   60 & 55 & b\\
    16 & CHEOPS & CH\_PR120053\_TG004602\_V0200       & 2022-07-11T12:50 &  7.0 &  236 &   60 & 56 & b\\
    17 & CHEOPS & CH\_PR120053\_TG005101\_V0200       & 2022-09-02T11:59 &  6.9 &  222 &   60 & 54 & b\\
     1 & ASTEP+ & TOI-1130.02\_20230602\_B            & 2023-06-02T20:24 &  6.2 &  248 &   60 & -  & b\\
     1 & ASTEP+ & TOI-1130.02\_20230602\_R            & 2023-06-02T20:24 &  6.4 &  785 &   20 & -  & b\\
     2 & ASTEP+ & TOI-1130.01\_20230623\_B            & 2023-06-23T12:18 &  5.3 &  135 &   60 & -  & c\\
     2 & ASTEP+ & TOI-1130.01\_20230623\_R            & 2023-06-23T12:13 &  5.4 &  685 &   20 & -  & c\\
     - & TESS   & s0067-0000000254113311-0261-a\_fast & 2023-07-01T03:30 & -    &  -   &   20 & -  & b,c\\
     3 & ASTEP+ & TOI-1130.01\_20230709\_B            & 2023-07-10T06:07 &  4.0 &  162 &   60 & -  & c\\
     3 & ASTEP+ & TOI-1130.01\_20230709\_R            & 2023-07-10T05:25 &  4.7 &  593 &   20 & -  & c\\
     4 & ASTEP+ & TOI-1130.01\_20230718\_B            & 2023-07-18T09:23 & 10.7 &  390 &   60 & -  & c\\
     4 & ASTEP+ & TOI-1130.01\_20230718\_R            & 2023-07-18T09:23 & 10.8 & 1292 &   20 & -  & c\\
     5 & ASTEP+ & TOI-1130.02\_20230721\_B            & 2023-07-21T13:30 &  8.1 &  349 &   60 & -  & b\\
     5 & ASTEP+ & TOI-1130.02\_20230721\_R            & 2023-07-21T13:30 &  8.2 & 1016 &   20 & -  & b\\
    \noalign{\smallskip}
    \hline 
\end{tabular}

%% file: stellar_parameters.tex
\begin{table}
    \centering
    \caption{\label{tab:spectroparams}Results from the spectroscopic stellar modelling of TOI-1130.}   
    \resizebox{\columnwidth}{!}{%
    \begin{tabular}{lcccc}
        \hline
        \noalign{\smallskip}
                                   & \multicolumn{2}{c}{This Work} &             &            \\
                                   & SpecMatch-Emp & SME           & H20$^{(a)}$ & K23$^{(b)}$ \\
        \noalign{\smallskip}
        \hline
        \noalign{\smallskip} 
        $T_\mathrm{eff}$~(K)       & $4256 \pm 70$   & $4360 \pm 108$           & $4250 \pm 67$             & $4350 \pm 60$   \\
        $\log g_\star$~(cgs)       & $4.65 \pm 0.12$ & $4.55 \pm 0.07$          & $4.60 \pm 0.02$           & $4.62 \pm 0.04$ \\
        $\mathrm{[Fe/H]}$~(dex)    & $0.17 \pm 0.09$ & $0.11 \pm 0.10$          & $>0.2$                    & $0.30 \pm 0.06$ \\
        $\mathrm{[Si/H]}$~(dex)    & \ldots        & $0.12 \pm 0.11$            & \ldots                    & \ldots \\
        $\mathrm{[Mg/H]}$~(dex)    & \ldots        & $0.13 \pm 0.12$            & \ldots                    & \ldots \\
        $V \sin i$~(km~s$^{-1}$)   & \ldots        & $2.5 \pm 0.9$              & $4.0 \pm 0.5$             & $\leq 3$ \\
        $M_\star\, (M_\sun)$       & \ldots        & $0.722_{-0.037}^{+0.042}$  & $0.684_{-0.017}^{+0.016}$ & $0.71 \pm 0.02$ \\
        $R_\star\, (R_\sun)$       & \ldots        & $0.697 \pm 0.011$          & $0.687 \pm 0.015$         & $0.68 \pm 0.02$ \\
        $\rho_\star\, (\rho_\sun)$ & \ldots        & $2.13 \pm 0.16$            & $2.11 \pm 0.15$           & $2.12 \pm 0.15$ \\
        $t_\star$~(Gyr)            & \ldots        & $5.4_{-4.9}^{+5.7}$        & $8.2_{-4.9}^{+3.8}$       & $3.2$--$5$\\
        \hline 
    \end{tabular}
    }
    \tablefoot{
        $^{(a)}$ \citet{Huang2020ApJ...892L...7H}; 
        $^{(b)}$ \citet{Korth2023A&A...675A.115K}.
    }
\end{table}

%% file: T0s_b.tex
\begin{tabular}{rrl}
    \hline\hline
    \noalign{\smallskip}
    \multicolumn{3}{l}{
    TOI-1130 b
    }\\
    \multicolumn{3}{l}{
    $T_\mathrm{ref} = 2392.8147\pm0.0002 (0.0126)\, \mathrm{BTJD_{TDB}}$
    }\\
    \multicolumn{3}{l}{
    $P_\mathrm{lin} = 4.078668\pm0.000002 (0.000097)$~days
    }\\
    \hline
    \noalign{\smallskip}
    $T_0$         & $\sigma_{T_0}$ & source \\
    $1658.73999$ & $0.00710$     & TESS-S13\\
    $1662.81104$ & $0.00306$     & TESS-S13\\
    $1666.88336$ & $0.00576$     & TESS-S13\\
    $1670.95086$ & $0.00478$     & TESS-S13\\
    $1675.01606$ & $0.00486$     & TESS-S13\\
    $1679.07242$ & $0.00529$     & TESS-S13\\
    $1731.94194$ & $0.00091$     &   LCO-SSO\\
    $2037.94623$ & $0.00303$     & TESS-S27\\
    $2042.00827$ & $0.00157$     & TESS-S27\\
    $2046.07458$ & $0.00191$     & TESS-S27\\
    $2050.14405$ & $0.00333$     & TESS-S27\\
    $2054.21920$ & $0.00178$     & TESS-S27\\
    $2058.29358$ & $0.00342$     & TESS-S27\\
    $2364.28998$ & $0.00177$     &   CHEOPS\\
    $2376.50951$ & $0.00078$     &   LCO-SAAO\\
    $2380.58414$ & $0.00589$     &   CHEOPS\\
    $2384.66071$ & $0.00081$     &   LCO-CTIO\\
    $2388.73347$ & $0.00086$     &   CHEOPS\\
    $2392.81083$ & $0.00194$     &   LCO-CTIO\\
    $2421.37739$ & $0.00395$     &   CHEOPS\\
    $2429.54946$ & $0.00092$     &   LCO-SAAO\\
    $2433.63585$ & $0.00075$     &   LCO-CTIO\\
    $2441.80478$ & $0.00052$     &   CHEOPS\\
    $2454.05512$ & $0.00204$     &   CHEOPS\\
    $2723.21209$ & $0.00068$     &   CHEOPS\\
    $2727.30096$ & $0.00636$     &   CHEOPS\\
    $2751.78779$ & $0.00109$     &   CHEOPS\\
    $2764.01177$ & $0.00097$     &   CHEOPS\\
    $2772.15663$ & $0.00057$     &   CHEOPS\\
    $2825.10022$ & $0.00068$     &   CHEOPS\\
    $3098.48272$ & $0.00147$     &  ASTEP+ R\\
    $3098.60441$ & $0.00935$     &  ASTEP+ B\\
    $3126.92521$ & $0.00192$     & TESS-S67\\
    $3130.99262$ & $0.00143$     & TESS-S67\\
    $3135.05644$ & $0.00261$     & TESS-S67\\
    $3143.19552$ & $0.00168$     & TESS-S67\\
    $3147.26878$ & $0.00374$     &  ASTEP+ R\\
    $3147.26951$ & $0.00093$     & TESS-S67\\
    $3147.27052$ & $0.00204$     &  ASTEP+ B\\
    \hline
\end{tabular}

%% file: T0s_c.tex
\begin{tabular}{rrl}
    \hline\hline
    \noalign{\smallskip}
    \multicolumn{3}{l}{
    TOI-1130 c
    }\\
    \multicolumn{3}{l}{
    $T_\mathrm{ref} = 2392.66995\pm0.00007 (0.00218)\, \mathrm{BTJD_{TDB}}$
    }\\
    \multicolumn{3}{l}{
    $P_\mathrm{lin} = 8.349560\pm0.000002 (0.000023)$~days
    }\\
    \hline
    \noalign{\smallskip}
    $T_0$         & $\sigma_{T_0}$ & source \\
    $1657.90383$ & $0.00103$ & TESS-S13\\
    $1666.25464$ & $0.00086$ & TESS-S13\\
    $1674.60532$ & $0.00094$ & TESS-S13\\
    $1758.10894$ & $0.00090$ &   PEST\\
    $1975.18610$ & $0.00037$ &   LCO-SSO\\
    $2008.58978$ & $0.00054$ &   LCO-SAAO\\
    $2041.99674$ & $0.00062$ & TESS-S27\\
    $2050.34671$ & $0.00035$ & TESS-S27\\
    $2058.69699$ & $0.00076$ & TESS-S27\\
    $2067.04526$ & $0.00029$ &   ASTEP, LCO-SSO, PEST\\
    $2083.74135$ & $0.00064$ &   CDK14\\
    $2092.08823$ & $0.00059$ &   PEST\\
    $2317.51746$ & $0.00052$ &   LCO-SAAO\\
    $2375.97590$ & $0.00026$ &   CHEOPS\\
    $2392.67431$ & $0.00028$ &   LCO-CTIO\\
    $2409.37117$ & $0.00029$ &   CHEOPS\\
    $2426.06660$ & $0.00037$ &   LCO-SSO\\
    $2434.41375$ & $0.00021$ &   CHEOPS\\
    $2451.10989$ & $0.00047$ &   ASTEP\\
    $2459.45902$ & $0.00018$ &   CHEOPS\\
    $2726.65265$ & $0.00025$ &   CHEOPS\\
    $2751.69770$ & $0.00021$ &   CHEOPS\\
    $3119.08278$ & $0.00035$ &  ASTEP+ R\\
    $3119.08399$ & $0.00148$ &  ASTEP+ B\\
    $3127.43479$ & $0.00059$ & TESS-S67\\
    $3135.78403$ & $0.00138$ &  ASTEP+ R\\
    $3135.78607$ & $0.00070$ & TESS-S67\\
    $3135.78661$ & $0.00233$ &  ASTEP+ B\\
    $3144.13645$ & $0.00043$ & TESS-S67\\
    $3144.13705$ & $0.00041$ &  ASTEP+ R\\
    $3144.13902$ & $0.00112$ &  ASTEP+ B\\
    \hline
\end{tabular}

%% file: phototrades_table.tex
\longtab{
\begin{longtable}{lcclccc}
\caption{\label{tab:phototrades} Fitted and physical parameters from the photo-dynamical model with \trades{}. 
Parameters as Maximum a Posteriori (MAP) and error as High Density Interval (HDI) at 68.27\%,
if the MAP is out-of HDI 68.27\% the HDI at 95.44\% is reported and indicated as $2\sigma$.
Last column shows the high-precision values of the principal parameters required for a precise N-body integration of the orbits.
Osculating parameters at reference time $t_\mathrm{epoch} = 2\,458\,657\, \mathrm{BJD_{TDB}}$.
}\\
\hline\hline
parameters & prior & MAP (HDI 68.27\%) & High-precision parameters\\
\hline
\endfirsthead
\caption{continued.}\\
\hline\hline
parameters & prior & MAP (HDI 68.27\%) & High-precision parameters\\
\hline
\endhead
\hline
\endfoot
stellar parameters &  & & \\
\textit{fitted} & & & \\
$R_\star\, (R_\sun)$         & $\mathcal{G}(0.697, 0.011)$ & $0.697_{-0.004}^{+0.005}$ & $0.697$\\
$\rho_\star\, (\rho_\sun)$   & $\mathcal{G}(2.13, 0.16)$   & $2.20_{-0.04}^{+0.04}$  & \\
LD~$q_{1,\mathrm{CHEOPS}}$   & $\mathcal{U}(0,1)$          & $0.50_{-0.03}^{+0.07}$  & \\
LD~$q_{2,\mathrm{CHEOPS}}$   & $\mathcal{U}(0,1)$          & $0.21_{-0.11}^{+0.06}$  & \\
LD~$q_{1,\mathrm{TESS}}$     & $\mathcal{U}(0,1)$          & $0.50_{-0.07}^{+0.05}$  & \\
LD~$q_{2,\mathrm{TESS}}$     & $\mathcal{U}(0,1)$          & $0.27_{-0.09}^{+0.12}$  & \\
LD~$q_{1,\mathrm{ASTEP+ B}}$ & $\mathcal{U}(0,1)$          & $0.38_{-0.05}^{+0.06}$  & \\
LD~$q_{2,\mathrm{ASTEP+ B}}$ & $\mathcal{U}(0,1)$          & $0.7_{-0.2}^{+0.1}$  & \\
LD~$q_{1,\mathrm{ASTEP+ R}}$ & $\mathcal{U}(0,1)$          & $0.1959_{-0.0010}^{+0.0859}$  & \\
LD~$q_{2,\mathrm{ASTEP+ R}}$ & $\mathcal{U}(0,1)$          & $0.8_{-0.5}^{+0.1}$  $(2\sigma)$ & \\
\textit{physical} & & & \\
$M_\star\, (M_\sun)$         & -                         & $0.745_{-0.009}^{+0.007}$ & $0.745$\\
LD~$u_{1,\mathrm{CHEOPS}}$   & $\mathcal{G}(0.53, 0.04)$ & $0.30_{-0.13}^{+0.08}$  & \\
LD~$u_{2,\mathrm{CHEOPS}}$   & $\mathcal{G}(0.12, 0.07)$ & $0.41_{-0.09}^{+0.17}$  & \\
LD~$u_{1,\mathrm{TESS}}$     & $\mathcal{G}(0.66, 0.05)$ & $0.4_{-0.1}^{+0.2}$  & \\
LD~$u_{2,\mathrm{TESS}}$     & $\mathcal{G}(0.06, 0.09)$ & $0.3_{-0.2}^{+0.1}$  & \\
LD~$u_{1,\mathrm{ASTEP+ B}}$ & $\mathcal{G}(0.37, 0.10)$ & $0.8_{-0.1}^{+0.2}$  & \\
LD~$u_{2,\mathrm{ASTEP+ B}}$ & $\mathcal{G}(0.25, 0.10)$ & $-0.2_{-0.2}^{+0.2}$  & \\
LD~$u_{1,\mathrm{ASTEP+ R}}$ & $\mathcal{G}(0.37, 0.10)$ & $0.70_{-0.27}^{+0.01}$  & \\
LD~$u_{2,\mathrm{ASTEP+ R}}$ & $\mathcal{G}(0.25, 0.10)$ & $-0.3_{-0.1}^{+0.5}$  $(2\sigma)$ & \\

\hline
planet b &  & & \\
\textit{fitted} & & & \\
$\log_{10}(M_\rmb/M_\star)$     & $\mathcal{U}(-6.1, -3.1)$       & $-4.097_{-0.003}^{+0.002}$  & \\
$R_\rmb/R_\star$                & $\mathcal{U}(0.00048, 1.71394)$ & $0.0480_{-0.0003}^{+0.0002}$  & \\
$P_\rmb$~(days)                 & $\mathcal{U}(3.074, 5.074)$     & $4.074554_{-0.000441}^{+0.000001}$  $(2\sigma)$ & $4.074554$\\
$\sqrt{e}\cos\omega_\rmb$       & $\mathcal{U}(-0.5, 0.5)$        & $-0.17777_{-0.00259}^{+0.00003}$  $(2\sigma)$ & \\
$\sqrt{e}\sin\omega_\rmb$       & $\mathcal{U}(-0.5, 0.5)$        & $0.1434_{-0.0003}^{+0.0005}$  & \\
$\lambda_\rmb\, (\degr)$\footnote{$\lambda$ is the mean longitude, given by the sum of
    argument of periastron $\omega$,
    mean anomaly $\mathcal{M}$,
    and longitude of ascending node $\Omega$.}
                                & $\mathcal{U}(0, 360)$   & $120.81_{-0.04}^{+0.13}$  $(2\sigma)$ & \\
$i_\rmb\, (\degr)$              & $\mathcal{U}(70, 120)$  & $87.49_{-0.08}^{+0.02}$  & $87.494901$\\
\textit{physical} & & & \\
$M_\rmb\, (M_\oplus)$           & -                      & $19.8_{-0.3}^{+0.2}$ & $19.833346$\\
$K_\rmb$~(\mps)            & -                      & $9.67_{-0.08}^{+0.05}$ & \\
$R_\rmb\, (R_\oplus)$           & -                      & $3.66_{-0.04}^{+0.03}$ & $3.657$\\
$\rho_\rmb\, (\rho_\oplus)$     & -                      & $0.41_{-0.01}^{+0.01}$ & \\
$\rho_\rmb$~(\gcc)          & -                      & $2.23_{-0.06}^{+0.06}$  & \\
$a_\rmb$~(au)                   & -                      & $0.0453_{-0.0002}^{+0.0001}$ & $0.045262$\\
$e_\rmb$                        & $\mathcal{U}(0, 0.25)$ & $0.052162_{-0.000002}^{+0.000956}$  $(2\sigma)$ & $0.052162$\\
$\omega_\rmb\, (\degr)$         & -                      & $141.11_{-0.09}^{+0.47}$  $(2\sigma)$ & $141.111127$\\
$\mathcal{M}_\rmb\, (\degr)$    & -                      & $159.7_{-0.4}^{+0.1}$  $(2\sigma)$ & $159.696701$\\
$\Omega_\rmb\, (\degr)$         & $180$~fixed            & - & \\
$T_\mathrm{eq,1,\rmb}$\footnote{
Equilibrium Temperature with 
$T_\mathrm{eq}$ with $A_B =0$ and $f = 1$.
}~(K)                           & -                      & $825_{-23}^{+23}$ & \\
TSM$_{1,\rmb}$\footnote{
Transmission Spectroscopic Metric from \citet{Kempton2018PASP..130k4401K} with 
$T_\mathrm{eq}$ with Bond albedo $A_B =0$ and $f = 1$.
}                               & -                      & $83_{-5}^{+5}$ & \\
$T_\mathrm{eq,2,\rmb}$\footnote{
Equilibrium Temperature $T_\mathrm{eq}$ with $A_B =0$ and $f = 2$.
}~(K)                           & -                      & $982_{-27}^{+28}$ & \\
TSM$_{2,\rmb}$\footnote{
Transmission Spectroscopic Metric (TSM) from \citet{Kempton2018PASP..130k4401K} with 
$T_\mathrm{eq}$ with $A_B =0$ and $f = 2$.
}                               & -                      & $98_{-6}^{+6}$ & \\

\hline
planet c &  & & \\
\textit{fitted} & & & \\
$\log_{10}(M_\rmc/M_\star)$     & $\mathcal{U}(-4.87, -1.87)$     & $-2.8687_{-0.0025}^{+0.0002}$  $(2\sigma)$ & \\
$R_\rmc/R_\star$                & $\mathcal{U}(0.00048, 1.71394)$ & $0.171_{-0.005}^{+0.005}$  & \\
$P_\rmc$~(days)                 & $\mathcal{U}(7.35, 9.35)$       & $8.3501898_{-0.0000010}^{+0.0000938}$  $(2\sigma)$ & $8.3501898$\\
$\sqrt{e}\cos\omega_\rmc$       & $\mathcal{U}(-0.5, 0.5)$        & $-0.1992_{-0.0022}^{+0.0004}$  $(2\sigma)$ & \\
$\sqrt{e}\sin\omega_\rmc$       & $\mathcal{U}(-0.5, 0.5)$        & $-0.009_{-0.001}^{+0.002}$  & \\
$\lambda_\rmc\, (\degr)$        & $\mathcal{U}(0, 360)$           & $235.56_{-0.04}^{+0.04}$  & \\
$i_\rmc\, (\degr)$              & $\mathcal{U}(70, 120)$          & $87.61_{-0.04}^{+0.04}$  & $87.613475$ \\
$\Omega_\rmc\, (\degr)$         & $\mathcal{U}(0, 360)$           & $179.99_{-0.10}^{+0.03}$  $(2\sigma)$ & $179.993043$\\
\textit{physical} & & & \\
$M_\rmc\, (M_\oplus)$           & -                      & $336_{-5}^{+2}$ & $335.603435$\\
$K_\rmc$~(\mps)            & -                      & $128.59_{-0.80}^{+0.01}$ & \\
$R_\rmc\, (R_\oplus)$           & -                      & $13.0_{-0.4}^{+0.4}$  & $12.983016$\\
$\rho_\rmc\, (\rho_\oplus)$     & -                      & $0.15_{-0.02}^{+0.01}$  & \\
$\rho_\rmc$~(\gcc)          & -                      & $0.84_{-0.09}^{+0.08}$  & \\
$a_\rmc$~(au)                   & -                      & $0.0731_{-0.0003}^{+0.0002}$  & $0.073056$\\
$e_\rmc$                        & $\mathcal{U}(0, 0.25)$ & $0.0398_{-0.0002}^{+0.0009}$  $(2\sigma)$ & $0.039773$\\
$\omega_\rmc\, (\degr)$         & -                      & $182.5_{-0.5}^{+0.4}$  & $182.502357$\\
$\mathcal{M}_\rmc\, (\degr)$    & -                      & $233.1_{-0.3}^{+0.5}$  & $233.068994$\\
$T_\mathrm{eq,1,\rmc}$~(K)      & -                      & $650_{-18}^{+18}$ & \\
TSM$_{1,\rmc}$                  & -                      & $135_{-17}^{+18}$ & \\
$T_\mathrm{eq,2,\rmc}$~(K)      & -                      & $773_{-21}^{+22}$ & \\
TSM$_{2,\rmc}$                  & -                      & $160_{-20}^{+21}$ & \\

\hline
radial velocities &  & & \\
\textit{fitted} & & & \\
$\gamma_\mathrm{CHIRON}$~(\mps)         & $\mathcal{U}(-10567, -8567)$ & $-9568_{-4}^{+4}$ &  \\
$\gamma_\mathrm{HARPS}$~(\mps)     & $\mathcal{U}(-9023, -7023)$  & $-8023_{-1}^{+3}$  & \\
$\gamma_\mathrm{PFS}$~(\mps)       & $\mathcal{U}(-960, -1040)$   & $39.8_{-0.9}^{+2.7}$  & \\
$\log_{2}\sigma_\mathrm{jitter,CHIRON}$ & $\mathcal{U}(-13.29, 6.64)$  & $0.1_{-10.3}^{+0.2}$  & \\
$\log_{2}\sigma_\mathrm{jitter,HARPS}$  & $\mathcal{U}(-13.29, 6.64)$  & $1.3_{-0.2}^{+0.2}$ &  \\
$\log_{2}\sigma_\mathrm{jitter,PFS}$    & $\mathcal{U}(-13.29, 6.64)$  & $1.31_{-0.05}^{+0.38}$ &  \\
RV linear trend~(\mps/d)      & $\mathcal{U}(-1, 1)$         & $0.486_{-0.026}^{+0.009}$  & \\
\textit{physical} & & & \\
$\sigma_\mathrm{jitter,CHIRON}$~(\mps) & -  & $1_{-1}^{+6}$  $(2\sigma)$ & \\
$\sigma_\mathrm{jitter,HARPS}$~(\mps)  & -  & $2.5_{-0.3}^{+0.3}$  & \\
$\sigma_\mathrm{jitter,PFS}$~(\mps)    & -  & $2.5_{-0.1}^{+0.7}$ &  \\

\hline
CHEOPS photometry &  & & \\
\textit{fitted} & & & \\
$f_{14}\, \mathrm{const.}$\footnote{$f$ stands for the normalised flux.} 
                           & $\mathcal{U}(0.5, 1.5)$  & $1.0043_{-0.0024}^{+0.0005}$  $(2\sigma)$ &  \\
$\rmd f_{14}/\rmd t$               & $\mathcal{U}(-1, 1)$     & $-0.0015_{-0.0004}^{+0.0004}$  & \\
$\rmd f_{14}/\rmd x$               & $\mathcal{U}(-1, 1)$     & $0.0008_{-0.0003}^{+0.0001}$  $(2\sigma)$ & \\ 
$\rmd ^{2}f_{14}/\rmd x^{2}$       & $\mathcal{U}(-1, 1)$     & $0.00070_{-0.00034}^{+0.00001}$  & \\
$\rmd f_{14}/\rmd y$               & $\mathcal{U}(-1, 1)$     & $-0.0005_{-0.0001}^{+0.0001}$  & \\
$\rmd ^{2}f_{14}/\rmd y^{2}$       & $\mathcal{U}(-1, 1)$     & $0.00034_{-0.00027}^{+0.00001}$  & \\
$\rmd ^{2}f_{14}/\rmd x \rmd y$         & $\mathcal{U}(-1, 1)$     & $-0.0005_{-0.0002}^{+0.0003}$  & \\
$\rmd f_{14}/\rmd bg$              & $\mathcal{U}(-1, 1)$     & $-0.00086_{-0.00006}^{+0.00072}$  & \\
$\rmd f_{14}/\rmd \cos\phi$\footnote{$\phi$ is the CHEOPS roll angle.} 
                           & $\mathcal{U}(-1, 1)$     & $0.0035_{-0.0036}^{+0.0008}$  $(2\sigma)$ & \\
$\rmd f_{14}/\rmd \sin\phi$        & $\mathcal{U}(-1, 1)$     & $-0.0014_{-0.0004}^{+0.0015}$  $(2\sigma)$ & \\
$\rmd f_{14}/\rmd \cos2\phi$       & $\mathcal{U}(-1, 1)$     & $0.0016_{-0.0016}^{+0.0003}$  $(2\sigma)$ & \\
$\rmd f_{14}/\rmd \sin2\phi$       & $\mathcal{U}(-1, 1)$     & $-0.0016_{-0.0004}^{+0.0016}$  $(2\sigma)$ & \\
$\rmd f_{14}/\rmd \cos3\phi$       & $\mathcal{U}(-1, 1)$     & $0.000233_{-0.000239}^{+0.000004}$  & \\
$\rmd f_{14}/\rmd \sin3\phi$       & $\mathcal{U}(-1, 1)$     & $-0.0007_{-0.0002}^{+0.0007}$  $(2\sigma)$ & \\

ASTEP+ B photometry &  & & \\
\textit{fitted} & & & \\
$f_{1}\, \mathrm{const.}$ & $\mathcal{U}(0.5, 1.5)$  & $1.0066_{-0.0008}^{+0.0002}$  & \\
$\rmd f_{1}/\rmd t$               & $\mathcal{U}(-1, 1)$     & $-0.062_{-0.003}^{+0.011}$ &  \\
$\rmd ^{2}f_{1}/\rmd t^{2}$       & $\mathcal{U}(-1, 1)$     & $0.161_{-0.039}^{+0.009}$ &  \\
$\rmd f_{1}/\rmd x$               & $\mathcal{U}(-1, 1)$     & $-0.00015_{-0.00007}^{+0.00006}$ &  \\
$\rmd f_{1}/\rmd y$               & $\mathcal{U}(-1, 1)$     & $0.0003_{-0.0002}^{+0.0001}$  & \\
$\rmd f_{1}/\rmd fwhm$            & $\mathcal{U}(-1, 1)$     & $0.00033_{-0.00003}^{+0.00022}$ &  \\
$\rmd f_{1}/\rmd sky$             & $\mathcal{U}(-1, 1)$     & $0.00013_{-0.00014}^{+0.00004}$  & \\

$f_{2}\, \mathrm{const.}$ & $\mathcal{U}(0.5, 1.5)$  & $1.0036_{-0.0007}^{+0.0002}$  & \\
$\rmd f_{2}/\rmd t$               & $\mathcal{U}(-1, 1)$     & $0.019_{-0.004}^{+0.015}$ &  \\
$\rmd ^{2}f_{2}/\rmd t^{2}$       & $\mathcal{U}(-1, 1)$     & $-0.13_{-0.07}^{+0.01}$ &  \\
$\rmd f_{2}/\rmd x$               & $\mathcal{U}(-1, 1)$     & $-0.000449_{-0.000078}^{+0.000002}$ &  \\
$\rmd f_{2}/\rmd y$               & $\mathcal{U}(-1, 1)$     & $-0.00031_{-0.00003}^{+0.00007}$ &  \\
$\rmd f_{2}/\rmd fwhm$            & $\mathcal{U}(-1, 1)$     & $-0.00032_{-0.00008}^{+0.00005}$ &  \\
$\rmd f_{2}/\rmd sky$             & $\mathcal{U}(-1, 1)$     & $0.0011_{-0.0002}^{+0.0001}$  & \\

$f_{3}\, \mathrm{const.}$ & $\mathcal{U}(0.5, 1.5)$  & $1.0004_{-0.0002}^{+0.0002}$ &  \\
$\rmd f_{3}/\rmd t$               & $\mathcal{U}(-1, 1)$     & $-0.009_{-0.002}^{+0.001}$ &  \\
$\rmd ^{2}f_{3}/\rmd t^{2}$       & $\mathcal{U}(-1, 1)$     & $0.037_{-0.002}^{+0.004}$ &  \\
$\rmd f_{3}/\rmd x$               & $\mathcal{U}(-1, 1)$     & $-0.00049_{-0.00001}^{+0.00012}$ &  \\
$\rmd f_{3}/\rmd y$               & $\mathcal{U}(-1, 1)$     & $-0.00101_{-0.00010}^{+0.00004}$ &  \\
$\rmd f_{3}/\rmd fwhm$            & $\mathcal{U}(-1, 1)$     & $0.00022_{-0.00007}^{+0.00003}$ &  \\
$\rmd f_{3}/\rmd sky$             & $\mathcal{U}(-1, 1)$     & $-0.00007_{-0.00005}^{+0.00007}$ &  \\

$f_{4}\, \mathrm{const.}$ & $\mathcal{U}(0.5, 1.5)$  & $0.9993_{-0.0003}^{+0.0002}$ &  \\
$\rmd f_{4}/\rmd t$               & $\mathcal{U}(-1, 1)$     & $0.012_{-0.005}^{+0.007}$ &  \\
$\rmd ^{2}f_{4}/\rmd t^{2}$       & $\mathcal{U}(-1, 1)$     & $-0.01_{-0.03}^{+0.03}$ &  \\
$\rmd f_{4}/\rmd x$               & $\mathcal{U}(-1, 1)$     & $0.00047_{-0.00015}^{+0.00003}$ &  \\
$\rmd f_{4}/\rmd y$               & $\mathcal{U}(-1, 1)$     & $0.000011_{-0.000165}^{+0.000008}$ &  \\
$\rmd f_{4}/\rmd fwhm$            & $\mathcal{U}(-1, 1)$     & $0.00011_{-0.00021}^{+0.00003}$ &  \\
$\rmd f_{4}/\rmd sky$             & $\mathcal{U}(-1, 1)$     & $-0.0011_{-0.0003}^{+0.0002}$  & \\

$f_{5}\, \mathrm{const.}$ & $\mathcal{U}(0.5, 1.5)$  & $0.9998_{-0.0001}^{+0.0004}$  $(2\sigma)$ & \\
$\rmd f_{5}/\rmd t$               & $\mathcal{U}(-1, 1)$     & $0.003_{-0.009}^{+0.002}$  $(2\sigma)$ & \\
$\rmd ^{2}f_{5}/\rmd t^{2}$       & $\mathcal{U}(-1, 1)$     & $0.003_{-0.008}^{+0.029}$  $(2\sigma)$ & \\
$\rmd f_{5}/\rmd x$               & $\mathcal{U}(-1, 1)$     & $0.00014_{-0.00003}^{+0.00005}$  & \\
$\rmd f_{5}/\rmd y$               & $\mathcal{U}(-1, 1)$     & $-0.00002_{-0.00013}^{+0.00003}$  $(2\sigma)$ & \\
$\rmd f_{5}/\rmd fwhm$            & $\mathcal{U}(-1, 1)$     & $0.000367_{-0.000008}^{+0.000067}$ &  \\
$\rmd f_{5}/\rmd sky$             & $\mathcal{U}(-1, 1)$     & $-0.00057_{-0.00030}^{+0.00007}$  $(2\sigma)$ & \\

ASTEP+ R photometry &  & & \\
\textit{fitted} & & & \\
$f_{1}\, \mathrm{const.}$ & $\mathcal{U}(0.5, 1.5)$  & $1.0053_{-0.0001}^{+0.0003}$ &  \\
$\rmd f_{1}/\rmd t$               & $\mathcal{U}(-1, 1)$     & $-0.036_{-0.005}^{+0.002}$ &  \\
$\rmd ^{2}f_{1}/\rmd t^{2}$       & $\mathcal{U}(-1, 1)$     & $0.15_{-0.01}^{+0.01}$ &  \\
$\rmd f_{1}/\rmd x$               & $\mathcal{U}(-1, 1)$     & $0.000213_{-0.000064}^{+0.000010}$ &  \\
$\rmd f_{1}/\rmd y$               & $\mathcal{U}(-1, 1)$     & $0.000146_{-0.000007}^{+0.000077}$ &  \\
$\rmd f_{1}/\rmd fwhm$            & $\mathcal{U}(-1, 1)$     & $0.00064_{-0.00020}^{+0.00005}$  $(2\sigma)$&  \\
$\rmd f_{1}/\rmd sky$             & $\mathcal{U}(-1, 1)$     & $-0.000080_{-0.000134}^{+0.000007}$  & \\

$f_{2}\, \mathrm{const.}$ & $\mathcal{U}(0.5, 1.5)$  & $1.0069_{-0.0005}^{+0.0002}$ &  \\
$\rmd f_{2}/\rmd t$               & $\mathcal{U}(-1, 1)$     & $-0.028_{-0.004}^{+0.011}$ &  \\
$\rmd ^{2}f_{2}/\rmd t^{2}$       & $\mathcal{U}(-1, 1)$     & $-0.00_{-0.05}^{+0.02}$ &  \\
$\rmd f_{2}/\rmd x$               & $\mathcal{U}(-1, 1)$     & $-0.00010_{-0.00007}^{+0.00004}$ &  \\
$\rmd f_{2}/\rmd y$               & $\mathcal{U}(-1, 1)$     & $-0.000080_{-0.000007}^{+0.000114}$ &  \\
$\rmd f_{2}/\rmd fwhm$            & $\mathcal{U}(-1, 1)$     & $0.000233_{-0.000147}^{+0.000007}$ &  \\
$\rmd f_{2}/\rmd sky$             & $\mathcal{U}(-1, 1)$     & $-0.00090_{-0.00023}^{+0.00007}$  & \\

$f_{3}\, \mathrm{const.}$ & $\mathcal{U}(0.5, 1.5)$  & $0.9949_{-0.0001}^{+0.0001}$ &  \\
$\rmd f_{3}/\rmd t$               & $\mathcal{U}(-1, 1)$     & $0.058_{-0.001}^{+0.001}$  & \\
$\rmd ^{2}f_{3}/\rmd t^{2}$       & $\mathcal{U}(-1, 1)$     & $-0.101_{-0.003}^{+0.002}$ &  \\
$\rmd f_{3}/\rmd x$               & $\mathcal{U}(-1, 1)$     & $-0.00038_{-0.00001}^{+0.00007}$  & \\
$\rmd f_{3}/\rmd y$               & $\mathcal{U}(-1, 1)$     & $0.00046_{-0.00004}^{+0.00005}$ &  \\
$\rmd f_{3}/\rmd fwhm$            & $\mathcal{U}(-1, 1)$     & $0.00017_{-0.00001}^{+0.00005}$ &  \\
$\rmd f_{3}/\rmd sky$             & $\mathcal{U}(-1, 1)$     & $0.000138_{-0.000004}^{+0.000059}$ &  \\

$f_{4}\, \mathrm{const.}$ & $\mathcal{U}(0.5, 1.5)$  & $0.9985_{-0.0002}^{+0.0007}$  $(2\sigma)$ &  \\
$\rmd f_{4}/\rmd t$               & $\mathcal{U}(-1, 1)$     & $0.048_{-0.016}^{+0.005}$  $(2\sigma)$ & \\
$\rmd ^{2}f_{4}/\rmd t^{2}$       & $\mathcal{U}(-1, 1)$     & $-0.183_{-0.001}^{+0.041}$ &  \\
$\rmd f_{4}/\rmd x$               & $\mathcal{U}(-1, 1)$     & $-0.00016_{-0.00002}^{+0.00023}$  $(2\sigma)$ & \\
$\rmd f_{4}/\rmd y$               & $\mathcal{U}(-1, 1)$     & $-0.00025_{-0.00004}^{+0.00011}$ &  \\
$\rmd f_{4}/\rmd fwhm$            & $\mathcal{U}(-1, 1)$     & $0.00001_{-0.00009}^{+0.00001}$ &  \\
$\rmd f_{4}/\rmd sky$             & $\mathcal{U}(-1, 1)$     & $0.0001_{-0.0002}^{+0.0001}$ &  \\

$f_{5}\, \mathrm{const.}$ & $\mathcal{U}(0.5, 1.5)$  & $0.99852_{-0.00037}^{+0.00006}$  & \\
$\rmd f_{5}/\rmd t$               & $\mathcal{U}(-1, 1)$     & $0.022_{-0.002}^{+0.005}$ &  \\
$\rmd ^{2}f_{5}/\rmd t^{2}$       & $\mathcal{U}(-1, 1)$     & $-0.045_{-0.014}^{+0.008}$ &  \\
$\rmd f_{5}/\rmd x$               & $\mathcal{U}(-1, 1)$     & $0.00019_{-0.00020}^{+0.00005}$  $(2\sigma)$ & \\
$\rmd f_{5}/\rmd y$               & $\mathcal{U}(-1, 1)$     & $-0.00011_{-0.00015}^{+0.00002}$ &  \\
$\rmd f_{5}/\rmd fwhm$            & $\mathcal{U}(-1, 1)$     & $-0.00022_{-0.00003}^{+0.00007}$ &  \\
$\rmd f_{5}/\rmd sky$             & $\mathcal{U}(-1, 1)$     & $0.00044_{-0.00002}^{+0.00014}$ &  \\

\end{longtable}
}

%% file: acknowledgements.tex
CHEOPS is an ESA mission in partnership with Switzerland with important contributions to the payload and the ground segment from Austria, Belgium, France, Germany, Hungary, Italy, Portugal, Spain, Sweden, and the United Kingdom. The CHEOPS Consortium would like to gratefully acknowledge the support received by all the agencies, offices, universities, and industries involved. Their flexibility and willingness to explore new approaches were essential to the success of this mission. CHEOPS data analysed in this article will be made available in the CHEOPS mission archive (\url{https://cheops.unige.ch/archive_browser/}). 
LBo, GBr, VNa, IPa, GPi, RRa, GSc, VSi, and TZi acknowledge support from CHEOPS ASI-INAF agreement n. 2019-29-HH.0. 
This work has been carried out within the framework of the NCCR PlanetS supported by the Swiss National Science Foundation under grants 51NF40\_182901 and 51NF40\_205606. AL acknowledges support of the Swiss National Science Foundation under grant number  TMSGI2\_211697. 
This work has been carried out within the framework of the NCCR PlanetS supported by the Swiss National Science Foundation under grants 51NF40\_182901 and 51NF40\_205606. 
This work uses data obtained with the ASTEP+ telescope, at Concordia Station in Antarctica. ASTEP+ benefited from the support of the French and Italian polar agencies IPEV and PNRA in the framework of the Concordia station program, from OCA, INSU, Idex UCAJEDI (ANR-15-IDEX-01) and ESA through the Science Faculty of the European Space Research and Technology Centre (ESTEC). The Birmingham contribution to ASTEP+ is supported by the European Union’s Horizon 2020 research and innovation programme (grant’s agreement no. 803193/BEBOP), and from the Science and Technology Facilities Council (STFC; grant no. ST/S00193X/1, and ST/W002582/1).
ABr was supported by the SNSA. 
ACC acknowledges support from STFC consolidated grant number ST/V000861/1, and UKSA grant number ST/X002217/1. 
MNG is the ESA CHEOPS Project Scientist and Mission Representative, and as such also responsible for the Guest Observers (GO) Programme. MNG does not relay proprietary information between the GO and Guaranteed Time Observation (GTO) Programmes, and does not decide on the definition and target selection of the GTO Programme. 
TWi acknowledges support from the UKSA and the University of Warwick. 
TZi acknowledges NVIDIA Academic Hardware Grant Program for the use of the Titan V GPU card and the Italian MUR Departments of Excellence grant 2023-2027 "Quantum Frontiers". 
DG gratefully acknowledges the financial support from the grant for internationalization (GAND\_GFI\_23\_01) provided by the University of Turin (Italy).
YAl acknowledges support from the Swiss National Science Foundation (SNSF) under grant 200020\_192038. 
We acknowledge financial support from the Agencia Estatal de Investigación of the Ministerio de Ciencia e Innovación MCIN/AEI/10.13039/501100011033 and the ERDF "A way of making Europe" through projects PID2019-107061GB-C61, PID2019-107061GB-C66, PID2021-125627OB-C31, and PID2021-125627OB-C32, from the Centre of Excellence "Severo Ochoa" award to the Instituto de Astrofísica de Canarias (CEX2019-000920-S), from the Centre of Excellence "María de Maeztu" award to the Institut de Ciències de l’Espai (CEX2020-001058-M), and from the Generalitat de Catalunya/CERCA programme. 
We acknowledge financial support from the Agencia Estatal de Investigación of the Ministerio de Ciencia e Innovación MCIN/AEI/10.13039/501100011033 and the ERDF "A way of making Europe" through projects PID2019-107061GB-C61, PID2019-107061GB-C66, PID2021-125627OB-C31, and PID2021-125627OB-C32, from the Centre of Excellence "Severo Ochoa" award to the Instituto de Astrofísica de Canarias (CEX2019-000920-S), from the Centre of Excellence "María de Maeztu" award to the Institut de Ciències de l’Espai (CEX2020-001058-M), and from the Generalitat de Catalunya/CERCA programme. 
S.C.C.B. acknowledges support from FCT through FCT contracts nr. IF/01312/2014/CP1215/CT0004. 
C.B. acknowledges support from the Swiss Space Office through the ESA PRODEX program. 
P.E.C. is funded by the Austrian Science Fund (FWF) Erwin Schroedinger Fellowship, program J4595-N. 
This project was supported by the CNES. 
The Belgian participation to CHEOPS has been supported by the Belgian Federal Science Policy Office (BELSPO) in the framework of the PRODEX Program, and by the University of Liège through an ARC grant for Concerted Research Actions financed by the Wallonia-Brussels Federation. 
L.D. thanks the Belgian Federal Science Policy Office (BELSPO) for the provision of financial support in the framework of the PRODEX Programme of the European Space Agency (ESA) under contract number 4000142531. 
This work was supported by FCT - Funda\c{c}\~{a}o para a Ci\^{e}ncia e a Tecnologia through national funds and by FEDER through COMPETE2020 through the research grants UIDB/04434/2020, UIDP/04434/2020, 2022.06962.PTDC. 
O.D.S.D. is supported in the form of work contract (DL 57/2016/CP1364/CT0004) funded by national funds through FCT. 
B.-O. D. acknowledges support from the Swiss State Secretariat for Education, Research and Innovation (SERI) under contract number MB22.00046. 
This project has received funding from the Swiss National Science Foundation for project 200021\_200726. It has also been carried out within the framework of the National Centre of Competence in Research PlanetS supported by the Swiss National Science Foundation under grant 51NF40\_205606. The authors acknowledge the financial support of the SNSF. 
MF and CMP gratefully acknowledge the support of the Swedish National Space Agency (DNR 65/19, 174/18). 
M.G. is an F.R.S.-FNRS Senior Research Associate. 
CHe acknowledges support from the European Union H2020-MSCA-ITN-2019 under Grant Agreement no. 860470 (CHAMELEON). 
SH gratefully acknowledges CNES funding through the grant 837319. 
KGI is the ESA CHEOPS Project Scientist and is responsible for the ESA CHEOPS Guest Observers Programme. She does not participate in, or contribute to, the definition of the Guaranteed Time Programme of the CHEOPS mission through which observations described in this paper have been taken, nor to any aspect of target selection for the programme. 
J.K. acknowledges the Swedish Research Council (VR: Etableringsbidrag 2017-04945), and the common acknowledgment A.C., A.D., B.E., K.G., and J.K. acknowledge their role as ESA-appointed CHEOPS Science Team Members. 
K.W.F.L. was supported by Deutsche Forschungsgemeinschaft grants RA714/14-1 within the DFG Schwerpunkt SPP 1992, Exploring the Diversity of Extrasolar Planets. 
This work was granted access to the HPC resources of MesoPSL financed by the Region Ile de France and the project Equip@Meso (reference ANR-10-EQPX-29-01) of the programme Investissements d'Avenir supervised by the Agence Nationale pour la Recherche. 
ML acknowledges support of the Swiss National Science Foundation under grant number PCEFP2\_194576. 
PM acknowledges support from STFC research grant number ST/R000638/1. 
This work was also partially supported by a grant from the Simons Foundation (PI Queloz, grant number 327127). 
NCSa acknowledges funding by the European Union (ERC, FIERCE, 101052347). Views and opinions expressed are however those of the author(s) only and do not necessarily reflect those of the European Union or the European Research Council. Neither the European Union nor the granting authority can be held responsible for them. 
A. S. acknowledges support from the Swiss Space Office through the ESA PRODEX program. 
S.G.S. acknowledge support from FCT through FCT contract nr. CEECIND/00826/2018 and POPH/FSE (EC). 
The Portuguese team thanks the Portuguese Space Agency for the provision of financial support in the framework of the PRODEX Programme of the European Space Agency (ESA) under contract number 4000142255. 
GyMSz acknowledges the support of the Hungarian National Research, Development and Innovation Office (NKFIH) grant K-125015, a a PRODEX Experiment Agreement No. 4000137122, the Lend\"ulet LP2018-7/2021 grant of the Hungarian Academy of Science and the support of the city of Szombathely. 
V.V.G. is an F.R.S-FNRS Research Associate. 
JV acknowledges support from the Swiss National Science Foundation (SNSF) under grant PZ00P2\_208945. 
NAW acknowledges UKSA grant ST/R004838/1.
A.H.M.J.T acknowledges receiving funding from the Science and Technology Facilities Council (STFC; grant n$^\circ$ ST/S00193X/1).